\documentclass[amsmath,amssymb,aps,pra,superscriptaddress,twocolumn
]{revtex4-2}

\usepackage{amssymb}
\usepackage{graphicx}% Include figure files
\usepackage{subcaption} % 关键宏包：支持子图
\usepackage{dcolumn}% Align table columns on the decimal point
\usepackage{bm}% bold math
\usepackage{amsmath}
\usepackage{romannum}
\usepackage[colorlinks,urlcolor=cyan,citecolor=blue,linkcolor=magenta]{hyperref}
\usepackage{ragged2e}

\begin{document}
\title{One-dimensional asymmetrically interacting quantum droplets in Bose-Bose mixtures}

\affiliation{Zhejiang Key Laboratory of Quantum State Control and Optical Field Manipulation, Department of Physics, Zhejiang Sci-Tech University, Hangzhou 310018, China}
\affiliation{Lanzhou Center for Theoretical Physics, Key Laboratory of Theoretical Physics of Gansu Province, Key Laboratory of Quantum Theory and Applications of MoE, Gansu Provincial Research Center for Basic Disciplines of Quantum Physics, Lanzhou University, Lanzhou 730000, China}
\affiliation{School of Physics, Zhengzhou University, Zhengzhou 450001, China}

\author{Huiyun Xiao}
\thanks{These authors contributed equally to this work}
\affiliation{Zhejiang Key Laboratory of Quantum State Control and Optical Field Manipulation, Department of Physics, Zhejiang Sci-Tech University, Hangzhou 310018, China}

\author{Xinran Zhang}
\thanks{These authors contributed equally to this work}
\affiliation{Zhejiang Key Laboratory of Quantum State Control and Optical Field Manipulation, Department of Physics, Zhejiang Sci-Tech University, Hangzhou 310018, China}

\author{Junli Liu}
\affiliation{Zhejiang Key Laboratory of Quantum State Control and Optical Field Manipulation, Department of Physics, Zhejiang Sci-Tech University, Hangzhou 310018, China}

\author{Xucong Du}
\affiliation{School of Physics, Zhengzhou University, Zhengzhou 450001, China}

\author{Xiao-Long Chen}\email{xiaolongchen@zstu.edu.cn}
\affiliation{Zhejiang Key Laboratory of Quantum State Control and Optical Field Manipulation, Department of Physics, Zhejiang Sci-Tech University, Hangzhou 310018, China}
\affiliation{Lanzhou Center for Theoretical Physics, Key Laboratory of Theoretical Physics of Gansu Province, Key Laboratory of Quantum Theory and Applications of MoE, Gansu Provincial Research Center for Basic Disciplines of Quantum Physics, Lanzhou University, Lanzhou 730000, China}

\author{Yunbo Zhang}
\email{ybzhang@zstu.edu.cn}
\affiliation{Zhejiang Key Laboratory of Quantum State Control and Optical Field Manipulation, Department of Physics, Zhejiang Sci-Tech University, Hangzhou 310018, China}

\date{\today}

\begin{abstract}
We theoretically investigate ground-state properties and collective excitations of one-dimensional quantum droplets in asymmetric Bose-Bose mixtures with unequal intraspin interactions. Using the extended Gross-Pitaevskii equation supported by variational, sum-rule, and linearization methods, we show that the intraspin interaction ratio substantially alters the droplet's density profile, driving a transition from Gaussian-like to flat-top shapes. By examining two experimentally relevant parameter regions, we analyze density profiles, radii, peak densities, and excitation spectra to distinguish quantum phases and to depict phase diagrams in the space of asymmetric interaction ratio and total atom number. We carefully study the frequencies of both well-known dipole and breathing modes and less-explored spin-dipole and spin-breathing modes. The breathing-mode frequency decreases monotonically with interaction ratio, approaching asymptotically the result of a conventional weakly interacting Bose gas. It varies nonmonotonically with total atom number, peaking at a critical point that highlights the crucial role of quantum fluctuations. In contrast, spin modes display distinct temporal spin density distributions and reveal in-phase and out-of-phase relative dynamics between components. Their frequencies depend instead monotonically on the interaction ratio and atom number. Our results provide a comprehensive understanding of asymmetric quantum droplets and link to experimentally accessible regimes in ultracold $^{39}$K atomic gases.
\end{abstract}

\maketitle

%%%%%%%%%%%%%%%%%%%%%%%%%%%%%%%%%%%%%%%%%%%%%%%%%%%%%%
%%%%%%%%%%%%%%%%%%%%%%%%%%%%%%%%%%%%%%%%%%%%%%%%%%%%%%
\section{INTRODUCTION}

Quantum droplets represent an emergent quantum state within ultracold atomic gases. Their existence challenges the conventional understanding that attractive Bose-Einstein condensates (BECs) in free space cannot support a stable solution at the mean-field (MF) level. However, recent works find that by including the three-body interactions or the quantum-fluctuation-induced Lee-Huang-Yang (LHY) energy beyond the MF level, the system could form the self-bound quantum droplets spontaneously without the need for an external confinement. It can exhibit distinctive macroscopic quantum properties and nonlinear dynamical behaviors~\cite{lhy1957eigenvalues,bulgac2002dilute,petrov2015quantum,petrov2016ultradilute,petrov2023beyond}.
After the innovative proposal of the stabilization mechanism, 
this novel dilute quantum-droplet state has been successfully observed first in BECs with anisotropic dipolar interactions~\cite{kadau2016observing,ferrier2016observation,chomaz2016quantum,schmitt2016self,tanzi2019observation,bottcher2019dilute}, and later in Bose-Bose mixtures with isotropic contact interactions~\cite{cabrera2018quantum,semeghini2018self,cheiney2018bright,derrico2019observation,ferioli2019collisions,skov2021observation,guo2021lhy,cavicchioli2025dynamical}. Similar stabilization mechanisms have been further identified in Bose-Fermi mixtures~\cite{debraj2019quantum,rakshit2019Sselfbound} and dipolar mixtures~\cite{smith2021quantum,bisset2021quantum}. Nowadays, the intriguing intrinsic properties render quantum droplets an ideal platform for investigating self-organization phenomena and nonequilibrium dynamics in many-body quantum systems. Meanwhile, the successful developments in these experiments of quantum droplets have also stimulated intensive and extensive investigations into this novel quantum state~\cite{kartashov2019frontiers,bottcher2021new,luo2021new,guo2021new}.

Bulgac proposed that three-body Efimov correlations can compromise the attractive mean-field two-body interactions to support the existence of self-bound dilute Bose and Fermi quantum droplets~\cite{bulgac2002dilute}. However, this scheme is very tough to achieve due to the severe three-body loss in the atomic experiments. Later in 2015, Petrov conducted groundbreaking research on two-component Bose gases within the framework beyond-mean-field theory~\cite{petrov2015quantum}. The crucial role of the higher-order LHY energy correction term induced by quantum fluctuation effects is revealed to stabilize the system against collapse and leads to the formation of a self-bound state. This stabilization mechanism is subsequently extended to low-dimensional systems~\cite{petrov2016ultradilute}, establishing a universal theoretical system for describing the ground state and dynamic properties of quantum droplets. These pioneering works stimulated fruitful experiments of quantum droplets in Bose-Bose mixtures to study liquid-to-gas transition~\cite{cabrera2018quantum,semeghini2018self,derrico2019observation}, soliton-to-droplet transition~\cite{cheiney2018bright}, multiple quantum droplets~\cite{cavicchioli2025dynamical}, expansion dynamics~\cite{semeghini2018self,guo2021lhy}, collisions~\cite{ferioli2019collisions}, and monopole oscillations~\cite{skov2021observation}. 
Meanwhile, a series of theoretical methods, including quantum Monte Carlo methods~\cite{cikojevic2018ultradilute,parisi2019liquid}, nonperturbative variational Euler-Lagrange method~\cite{staudinger2018selfbound}, bosonic pairing theory~\cite{hu2020consistent}, higher-order Beliaev's approach~\cite{gu2020phonon}, Gaussian state theory~\cite{pan2022quantum}, and thermodynamic approach~\cite{he2023quantum}, have been developed to complementarily explore the properties of this novel self-bound state in binary Bose mixtures. These theoretical approaches stimulated extensive and intensive investigations of quantum droplets in bosonic mixtures, such as studies of soliton-droplet transition~\cite{cappellaro2018collective} or liquid-gas coexistence~\cite{spada2023attractive}, novel vortex quantum droplets~\cite{kartashov2018three,li2018two,kartashov2019metastability,otajonov2020variational},
effect of thermal fluctuations~\cite{guebli2021quantum}, dimensional crossover~\cite{zin2018quantum}, collisional or rotational properties~\cite{cikojevic2021dynamics,mithun2021statistical,hu2022collisional,tengstrand2022droplet,nikolaou2023rotating}, low-energy spectrum or collective excitations~\cite{astrakharchik2018dynamics,otajonov2019stationary,sturmer2021breathing,tylutki2020collective,hu2020collective,dong2022internal,du2023ground,fei2024collective,charalampidis2025twocomponent}, etc. However, most existing studies adopt symmetric intraspin interactions (i.e., $g_{1}=g_{2}$ with balanced atomic mixtures) to describe the quantum-droplet state using a scalar wave function, leaving the role of interaction asymmetry largely unexplored~\cite{flynn2023quantum,flynn2024harmonically,kartashov2024multipole}, which is closely associated to the current experiments of quantum droplets in Bose mixtures~\cite{cabrera2018quantum,semeghini2018self,cheiney2018bright,ferioli2019collisions,derrico2019observation,guo2021lhy,skov2021observation,cavicchioli2025dynamical}.
Furthermore, these works focus primarily on conventional collective modes, with minimal attention given to less-studied spin-dependent collective modes~\cite{li2012sumrules,sartori2015spindipole,bienaim2016soc,fava2018observation}.

In this work, we investigate the ground-state properties and collective excitation modes of one-dimensional quantum droplets in weakly interacting binary Bose gases with asymmetric intraspin interactions (i.e., $g_1\neq g_2$), by employing both numerical and analytical approaches as described in our previous works~\cite{hu2022collisional,du2023ground,fei2024collective}. The ground-state wave function is obtained by numerically solving the static extended Gross-Pitaevskii equation (GPE) for quantum droplets. In terms of this ground-state wave function, we systematically examine the density profile, cloud size, and peak density of the quantum droplets under varying interaction-strength ratio (i.e., $\lambda\equiv g_1/g_2$), trapping strength, and total atom number. The numerical results are compared with those obtained from a variational approach with a super-Gaussian ansatz for the droplet wave function. Furthermore, we introduce controlled perturbations to excite collective excitations in trapped droplets, and adopt the time‑dependent extended GPE to describe the behaviors of specific collective modes, such as the well-studied dipole, monopole (breathing), as well as the less-studied spin‑dipole and spin‑monopole (spin‑breathing) modes. Additionally, based on the super‑Gaussian variational ansatz, we derive analytical expressions for the dipole‑mode and breathing‑mode frequencies within a sum‑rule formalism to verify the numerical predictions. Meanwhile, we further apply the linearization technique and identify these four mode frequencies to justify the numerical results obtained from the time‑dependent extended GPE.

The remainder of this paper is organized as follows. Section~\ref{sec: theo} begins with the energy functional for quantum droplets in a one-dimensional weakly interacting two-component Bose mixture. We then derive the extended Gross-Pitaevskii equations with an external harmonic trapping potential to describe this quantum droplet. Meanwhile, a variational approximation with a super-Gaussian ansatz for the droplet wave function, the sum-rule approach, and the linearization technique are introduced successively. In Sec.~\ref{sec: results}, we apply these methods to consider the roles of interaction-strength ratio, trapping strength, and total atom number on droplets, and compute the ground-state properties as well as the low‑lying collective modes under asymmetric intraspin interactions. The influence of the interaction-strength ratio, trapping strength, and total atom number on the quantum droplets is systematically discussed. Finally, Sec.~\ref{sec: conclusions} provides concluding remarks and an outlook for future work.

%%%%%%%%%%%%%%%%%%%%%%%%%%%%%%%%%%%%%%%%%%%%%%%%%%%%%%
%%%%%%%%%%%%%%%%%%%%%%%%%%%%%%%%%%%%%%%%%%%%%%%%%%%%%%
\section{Theoretical framework} \label{sec: theo}

In this section, we present three theoretical methods for investigating the ground-state properties and collective excitations of one-dimensional quantum droplets in weakly interacting Bose-Bose mixtures, both qualitatively and quantitatively. We numerically solve the extended GPE to provide a comprehensive study of the static and dynamic behaviors, which will be compared with predictions from a variational approach, the sum-rule approach, and the linearization technique.

%%%%%%%%%%%%%%%%%%%%%%%%%%%%%%%%%%%%%%%%%%%%%%%%%%%%%%
\subsection{Time-dependent extended Gross-Pitaevskii equation}

We consider the quantum droplet in a one-dimensional (1D) weakly interacting Bose-Bose mixture with equal masses (i.e., $m=m_1=m_2$) stabilized by the beyond-MF or LHY energies arising from quantum fluctuations~\cite{petrov2015quantum}. This quantum droplet can be described by the 1D effective energy density functional including the LHY energy corrections~\cite{petrov2016ultradilute},
\begin{eqnarray} \label{eq:energyfunctional}
    E^\mathrm{(1D)}/V &=& \frac{(n_1 \sqrt{g_1} -n_2 \sqrt{g_2})^2}{2} + \frac{g\delta g(n_1 \sqrt{g_2} +n_2 \sqrt{g_1})^2}{(g_{1}+g_2)^2} \nonumber 
    \\
    &&- \frac{2\sqrt{m}(n_1 g_1 +n_2 g_2)^{3/2}}{3\pi \hbar},
\end{eqnarray}
near the collapsing point $0<\delta g\equiv g+g_{12}\ll g\equiv\sqrt{g_1g_2}$ of the mean-field level. Here, the first two terms at the right side of Eq.~\eqref{eq:energyfunctional} correspond to mean-field energy and the third term is the LHY energy. Note that the LHY energy functional possesses distinct expressions for different dimensions~\cite{petrov2015quantum,petrov2016ultradilute}. $g_{1,2}$ and $g_{12}$ are the effective 1D intra- and interspin interaction strengths, which are repulsive and attractive in this quantum droplet regime, respectively, i.e., $g_{1,2}>0$ and $g_{12}<0$. $n_{1,2}$ represents the density of two spin components.

In this work, instead of considering the symmetric interaction (i.e., $g_1=g_2$) as in most previous works, we emphasize the effect of asymmetric intraspin interactions with $g_1\neq g_2$ on quantum droplets and introduce a dimensionless interaction ratio, 
\begin{equation}
    \lambda\equiv\frac{g_1}{g_2},
\end{equation}
to quantify this asymmetry. For this purpose, by starting from the energy functional in Eq.~\eqref{eq:energyfunctional} and employing the Heisenberg equation of motion, we derive the extended GPE for an asymmetric 1D Bose-Bose mixture in an external harmonic trap~\cite{petrov2016ultradilute,mithun2020modulational}, 
\begin{eqnarray} \label{eq:eGPE}
&&i\hbar \frac{\partial}{\partial t} \psi_\sigma = \left[ H_\mathrm{ho} +\left(g_\sigma+\frac{2gg_{\bar{\sigma}}\delta g}{(g_1+g_2)^2} \right)n_\sigma  
  \nonumber \right.\\
 &&\left.-\left(g-\frac{2g^2\delta g}{(g_1+g_2)^2}\right) n_{\bar{\sigma}}-\frac{\sqrt{m}}{\pi \hbar}g_\sigma(g_1n_1  +g_2n_{2})^{\frac{1}{2}} \right] \psi_\sigma,
\end{eqnarray}
for spin component $\sigma=\{1,2\}$ with different spin indices $\sigma\neq\bar{\sigma}$. $H_\mathrm{ho}=-\frac{\hbar^2}{2m}\partial_x^2 +V_\mathrm{ext}(x)$ denotes the single-particle Hamiltonian of a harmonic oscillator, and the frequency $\omega_x=\kappa\omega_0$ of the trapping potential $V_\mathrm{ext}(x)=\frac{1}{2}m\omega_x^2 x^2$ can be tuned by a parameter $\kappa$. $n_\sigma\equiv\left|\psi_\sigma\right|^2$ represents the local density of two spin components, and the wave functions are normalized to $N$, i.e., $\sum_{\sigma=1,2}\int \left|\psi_\sigma\right|^2 dx=N$. 
In practical calculations, we adopt a set of characteristic units related to the harmonic oscillator, i.e., the characteristic energy $E_0\equiv\hbar\omega_0$ and length $x_0=\sqrt{\hbar/(m\omega_0)}$, to obtain dimensionless extended GPE. By separating the variables via $\psi_\sigma(x,t)=\psi_\sigma(x)e^{-i\mu t}$ with the chemical potential $\mu$ and numerically solving the time-independent coupled equations, we can study the static properties of quantum droplets. 

To further investigate the crucial low-energy collective excitations or elementary excitations of this quantum many-body system, we can slightly perturb the static system to excite it and probe its responses as in realistic cold-atomic experiments.
In general, we consider four typical low-energy collective excitations here, i.e., the familiar dipole and breathing (or monopole) modes, and the spin-dipole and spin-breathing modes~\cite{li2012sumrules, sartori2015spindipole,bienaim2016soc, fava2018observation}, by developing distinct modulations of the external trapping potential and studying the corresponding dynamical behaviors of the system via the time-dependent extended GPE in Eq.~\eqref{eq:eGPE}. In practice, we excite these four collective modes by abruptly changing the external potential $V_\mathrm{ext}(x)$ to a slightly changed potential $V'_\mathrm{ext}(x)$ as in our previous works~\cite{du2023ground,fei2024collective}. The specific form is given as
\begin{equation} \label{eq:pertubation}
V'_\mathrm{ext}(x) = 
    \begin{cases}
    \frac{1}{2}m \omega_x^2  (x-\mathcal{I}\delta x)^2\\
    \frac{1}{2}m (1+\mathcal{I}\chi)^2\omega_x^2 x^2\\
    \frac{1}{2}m \omega_x^2  (x-\sigma_z\delta x)^2\\
    \frac{1}{2}m (1+\sigma_z\chi)^2\omega_x^2 x^2,
    \end{cases}  
\end{equation}
for four respective collective modes. Here, $\delta x$ is a small displacement and $\chi\ll1$ is a small factor to tune the trapping frequency. $\mathcal{I}$ and $\sigma_z$ denote the $2\times2$ identity and Pauli matrices, respectively. 

For instance, by applying the same displacement $\mathcal{I}\delta x$ for both spin components as described by the first term in Eq.~\eqref{eq:pertubation}, the dipole mode is excited with the total density profile oscillating about the origin, which can be characterized by the time-dependent center-of-mass displacement $\left<x(t)\right>$ averaged on the time-dependent wave function.
Similarly, we tune the trapping frequency with the same factor $\mathcal{I}\chi$ for both spin components as the second term in Eq.~\eqref{eq:pertubation} and excite the monopole- or breathing-mode oscillation. In this mode, the atomic cloud will expand and contract monopolarly, behaving as breathing, which can be characterized by the dynamical evolution of the cloud size or the squared size $\left<x^2(t)\right>$. 
In contrast, we employ spin-dependent displacements $\sigma_z x$ or trapping frequency variation $\sigma_z\chi$ as in the last two terms in Eq.~\eqref{eq:pertubation} to excite the spin-dipole and spin-breathing modes, respectively. These two spin-related excitations reveal the relative motions of the two spin components, i.e., two spin densities counterpropagating along the $x$ axis in the spin-dipole mode or exhibiting relative trends of expansion and contraction in the spin-breathing mode. They can be further characterized by the corresponding displacement $\left< \sigma_z x(t) \right>$  and the squared size $\left< \sigma_z x^2(t) \right>$ of the spin density, respectively. 
Eventually, after slightly perturbing the system to excite respective collective modes, we calculate the time-dependent wave functions as well as the evolution of the corresponding physical observables $\left<\mathcal{O}(t)\right>$ by numerically solving the time-dependent extended GPE in Eq.~\eqref{eq:eGPE}. The mode frequency is further extracted from the frequency peaks in the Fourier analysis of $\left<\mathcal{O}(t)\right>$ and the spatiotemporal evolution of the density profile possessing characteristic oscillation patterns.

%%%%%%%%%%%%%%%%%%%%%%%%%%%%%%%%%%%%%%%%%%%%%%%%%%%%%%
\subsection{Variational approximation and the sum-rule approach}

In addition to the extended GPE, we further introduce a variational approximation to provide a qualitative and quantitative description of the static properties of quantum droplets in this asymmetric intraspin-interaction configuration. Meanwhile, the sum-rule approach is adopted to derive an explicit analytic expression of the dipole- and breathing-mode frequencies of quantum droplets that are compared to the numerical results. As studied in previous works~\cite{astrakharchik2018dynamics,otajonov2019stationary,tylutki2020collective,hu2020collective,sturmer2021breathing,dong2022internal,du2023ground,fei2024collective}, the quantum-fluctuation-stabilized droplets exhibit a unique flat-top structure in the density distribution at relatively large atom number in sharp contrast to the conventional Gaussian state. Therefore, one can usually employ the so-called super-Gaussian variational \emph{ansatz} to describe the spinor wave function of self-bound quantum droplets~\cite{otajonov2019stationary,otajonov2020variational,sturmer2021breathing,du2023ground},
\begin{subequations}\label{eq:ansatz}
    \begin{eqnarray}
     \psi_1^{SG}(x) &=& \sqrt{\frac{N\alpha}{\Gamma(\frac{1}{2\alpha})(\sqrt{\lambda}+1)\sigma}}e^{-\frac{1}{2}\left(\frac{x}{\sigma}\right)^{2\alpha}},
     \\
    \psi_2^{SG}(x) &=& \sqrt{\frac{N\alpha\sqrt{\lambda}}{\Gamma(\frac{1}{2\alpha})(\sqrt{\lambda}+1)\sigma}} e^{-\frac{1}{2}(\frac{x}{\sigma})^{2\alpha}},   
\end{eqnarray}
\end{subequations}
with two variables, i.e., the cloud width $\sigma$ and the exponent factor $\alpha$~\footnote{Note that we have tried the variational \emph{Ansatz} with four variables, i.e., two widths $\sigma_{1,2}$, and exponent factors $\alpha_{1,2}$. And after the minimization, they are equal ($\sigma_{1}=\sigma_{2}$ and $\alpha_{1}=\alpha_{2}$) for each spin.}. Here, $\Gamma(x)$ is the Gamma function. The width determines the size of the droplet, while the exponent factor reveals the Gaussian or flat-top nature in the density profile. In practice, we substitute the variational ansatz given by Eq.~\eqref{eq:ansatz} into the total energy functional including Eq.~\eqref{eq:energyfunctional} as well as the kinetic and external trapping energies, and minimize the energy to obtain the extrema across the parameter spaces. The explicit equations of the extremum points, as well as their behaviors, can be seen in detail in Appendix~\ref{app:VA-SuperG}. With the calculated extrema, the wave function of quantum droplets can be constructed to study the static properties. Furthermore, the sum-rule approach is employed to provide an approximate upper bound for the frequencies of relevant collective excitations~\cite{stringari1996collective,menotti2002collective,dalfovo1999theory,pitaevskii2016bose}. One can introduce the energy-weighted moments and relate them to the commutation relations of specific operators. Thus, the frequencies of the collective excitation can be estimated by the moment ratios. For instance, by introducing the excitation operator of the dipole mode $\hat{F} = \sum_{i=1}^N\hat{x}_i$, the frequency of the dipole mode is bounded by the moment ratio $m_3/m_1$ with $m_1=\frac{1}{2}\langle \psi|[\hat{F}^\dagger,[\hat{H},\hat{F}]]|\psi\rangle$ and $m_3=\frac{1}{2}\langle \psi|[[\hat{F}^\dagger,\hat{H}],[\hat{H},[\hat{H},\hat{F}]]]|\psi\rangle$~\cite{dalfovo1999theory,pitaevskii2016bose}. 
In the framework of first quantization, the Hamiltonian of a weakly interacting Bose mixture with $m_1=m_2=m$ is given as $\hat{H}=\sum_{i,\sigma}[-\frac{\hbar^2\nabla_{i,\sigma}^2}{2m}+V_\sigma(x_i)]+\frac{1}{2}\sum_{i,j,\sigma,\sigma\prime}U_{\sigma,\sigma\prime}(x_i-x_j)$ with $i,j=\{1,2,\ldots,N\}$ and $\sigma,\sigma\prime=\{1,2\}$. $V(x) = \frac{1}{2} m \omega_{x}^2 x^2$ represents the external potential in which the particle is placed and $U_{\sigma,\sigma\prime}(x_i - x_j)$ describes the contact pseudopotentials between any two particles.
This leads to the expression
\begin{equation} \label{eq:omega_d}
\omega_\mathrm{d}^2 = \frac{m_3}{m_1 \hbar^2} =\omega_x^2,
\end{equation}
which verifies the Kohn theorem that the oscillation frequency of the dipole mode equals the frequency of the trapping potential. In addition, the frequency of the breathing mode can be calculated by the moment ratio $\omega_\mathrm{b}^2=\frac{m_1}{m_{-1}\hbar^2}$ of the energy-weighted moment $m_1$ to the inverse-energy-weighted moment $m_{-1}$ with the excitation operator $\hat{F}=\sum_{i=1}^N\hat{x}_i^2$. Here, one can instead use the relation between the inverse energy-weighted moment and the static polarization $m_{-1} = (1/2)\chi$ with $\chi=-\frac{2N}{m}\frac{\partial\langle x^2\rangle}{\partial \omega_x^2}$ and obtains the useful result of the breathing-mode frequency $\omega_\mathrm{b}^2=\frac{m_1}{m_{-1}\hbar^2}=-\sigma\frac{\partial \omega_x^2}{\partial\sigma}$~\cite{menotti2002collective,hu2020collective,fei2024collective}. After some straightforward algebra, we derive an analytic expression for the breathing-mode frequency $\omega_\mathrm{b}$ in terms of the extremum points $(\alpha_0,\sigma_0)$ as
\begin{eqnarray} \label{eq:omega_b}
    \omega_\mathrm{b}^2&=&4\frac{\hbar^2\alpha_0^2\Gamma(2-\frac{1}{2\alpha_0})}{m^2\sigma_0^4\Gamma(\frac{3}{2\alpha_0})}+3\left(\frac{1}{2}\right)^{\frac{1}{2\alpha_0}}\frac{N\alpha_0 \sqrt{\lambda}(g_{12}+g_2\sqrt{\lambda})}{m\sigma_0^3\Gamma(\frac{3}{2\alpha_0})(\sqrt{\lambda}+1)^2}
   \nonumber\\
    && -
  \frac{5}{6}\left(\frac{2}{3}\right)^{\frac{1}{2\alpha_0}}\sqrt{\frac{N\alpha_0\Gamma(\frac{1}{2\alpha_0})}{\sigma_0^5}}\frac{(g_2\sqrt{\lambda})^{\frac{3}{2}}}{\sqrt{m}\pi\hbar\Gamma(\frac{3}{2\alpha_0})}.
\end{eqnarray}
Therefore, after calculating the extremum points $(\alpha_0,\sigma_0)$ in the minimization, the corresponding frequencies of the collective excitations, $\omega_\mathrm{d,b}$, can be constructed to be compared with the results from the extended GPE.

%%%%%%%%%%%%%%%%%%%%%%%%%%%%%%%%%%%%%%%%%%%%%%%%%
\subsection{Linearization technique} \label{sec:linearization}

To justify our analytical and numerical predictions presented in the previous two sections, we further introduce the linearization technique to calculate the collective excitation spectrum that is widely employed for quantum droplets~\cite{tylutki2020collective,hu2020collective,dong2022internal,du2023ground,fei2024collective,charalampidis2025twocomponent}. Following the standard procedure, we linearize the Bose field for spin component $\sigma=\{1,2\}$ to a combination of a static classic field and a time-dependent small-amplitude oscillation as
\begin{eqnarray}
    \Psi_\sigma(x,t) = e^{-i\mu t}\left\{\psi_\sigma(x) +\eta_\sigma(x,t)\right\},
\end{eqnarray}
with the chemical potential $\mu$, and expand the oscillation as $\eta_\sigma(x,t) = \sum_j \left[u_{j,\sigma}(x) e^{-i\omega_j t}+v^*_{j,\sigma}(x) e^{i\omega_j t}\right]$ in terms of the quasiparticle wave-function basis $(u_{j,\sigma},v_{j,\sigma})$, with the $j$th quasiparticle frequency $\omega_j$. After substituting this linearized field back into the extended GPE~\eqref{eq:eGPE}, one obtains the static extended GPE for the ground-state wave functions $\psi_\sigma(x)$ and the coupled equations for the oscillations to the first order of the small fluctuations $u_{j,\sigma},v_{j,\sigma}$. The coupled equations for the $j$th quasiparticle mode with frequency $\omega_j$ are given explicitly as
\begin{eqnarray}\label{eq:Linearization}
\begin{pmatrix}    \mathcal{L}_1 &-\mathcal{A} &\mathcal{B}_1 &-\mathcal{A} \\   -\mathcal{A} &\mathcal{L}_2 &-\mathcal{A} & \mathcal{B}_2\\  
-\mathcal{B}_1 &\mathcal{A} &-\mathcal{L}_1&\mathcal{A} \\
\mathcal{A} &-\mathcal{B}_2 &\mathcal{A} &-\mathcal{L}_2 \end{pmatrix}
\begin{pmatrix} u_1\\u_2\\    v_1\\v_2\end{pmatrix} = \omega_j
\begin{pmatrix}   u_1\\u_2\\    v_1\\v_2\end{pmatrix},
\end{eqnarray}
with the introduced matrix elements defined as 
\begin{subequations}
\begin{eqnarray}
    \mathcal{L_\sigma}&=& H_\mathrm{ho}-\mu+\left[g_\sigma+\frac{2gg_{\bar{\sigma}}\delta g}{(g_1+g_2)^2} \right]2n_\sigma -\left[1-\frac{2g\delta g}{(g_1+g_2)^2}\right]gn_{\bar{\sigma}}\nonumber\\
   & -&\frac{1}{2\pi} \frac{g^2_\sigma n_\sigma}{(n_1 g_1 +n_2 g_2)^{1/2}}-\frac{1}{\pi} g_\sigma (n_1 g_1 +n_2 g_2)^{1/2},
\end{eqnarray}
\begin{eqnarray}
    \mathcal{A}=\left[1-\frac{2g\delta g}{(g_1+g_2)^2}\right]g\sqrt{n_1 n_2}+\frac{1}{2\pi} \frac{g^2 \sqrt{n_1 n_2} }{(n_1 g_1 +n_2 g_2)^{1/2}},
\end{eqnarray}
\begin{eqnarray}
    \mathcal{B_\sigma}=\left[g_\sigma+\frac{2gg_{\bar{\sigma}}\delta g}{(g_1+g_2)^2} \right]n_\sigma - \frac{1}{2\pi } \frac{g^2_\sigma n_\sigma}{(n_1 g_1 +n_2 g_2)^{1/2}}.
\end{eqnarray}
\end{subequations}

Therefore, after solving the static extended GPE for the ground-state wave functions $\psi_\sigma(x)$, we can further solve the coupled equations in Eq.~\eqref{eq:Linearization} to calculate the corresponding quasiparticle wave functions $u_{j,\sigma},v_{j,\sigma}$ and frequencies $\omega_j$. Using these low-energy collective excitation frequencies, we can quantitatively compare and justify the oscillation frequencies of specific collective modes obtained from the sum-rule approach within the variational approximation and from the numerically solved time-dependent extended GPE.
\begin{figure}[ht]
\centering
\includegraphics[width=0.45\textwidth]  {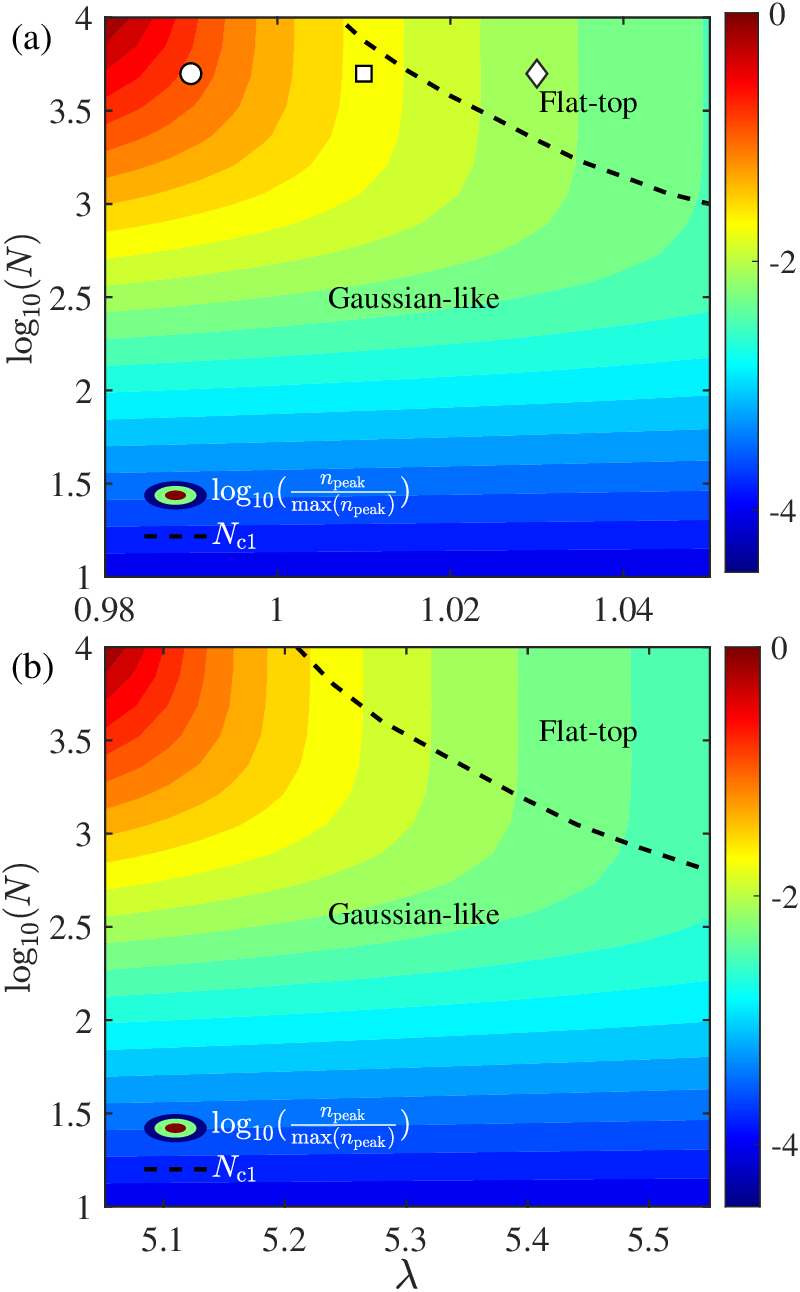}
\caption{\justifying
Typical phase diagram in the parameter space of interaction strength ratio $\lambda$ and total atom number $N$. (a), (b) The contour plots of the peak density of quantum droplets in the $\lambda$-$\log_{10}(N)$ plane at the parameter regions I and II described in the main text, respectively. The gradient color represents the scaled peak density $\log_{10}[n_\mathrm{peak}/\max(n_\mathrm{peak})]$. The dashed line marks the critical position $N_{c1}$ to distinguish the phase boundary between Gaussian-like and flat-top phases, corresponding to the saturation point of the peak density. Here, the hollow circle, square, and diamond in (a) denote the positions of three typical density profiles in Fig.~\ref{fig:densityprofile}(a). The interaction parameters in (a) region I and (b) region II are given as $g_2=0.100,g_{12}=-0.099$, and $g_2=0.089,g_{12}=-0.200$, respectively.} 
\label{fig:phasediagram}
\end{figure}
\begin{figure}[ht]
\centering
\includegraphics[width=0.45\textwidth]  {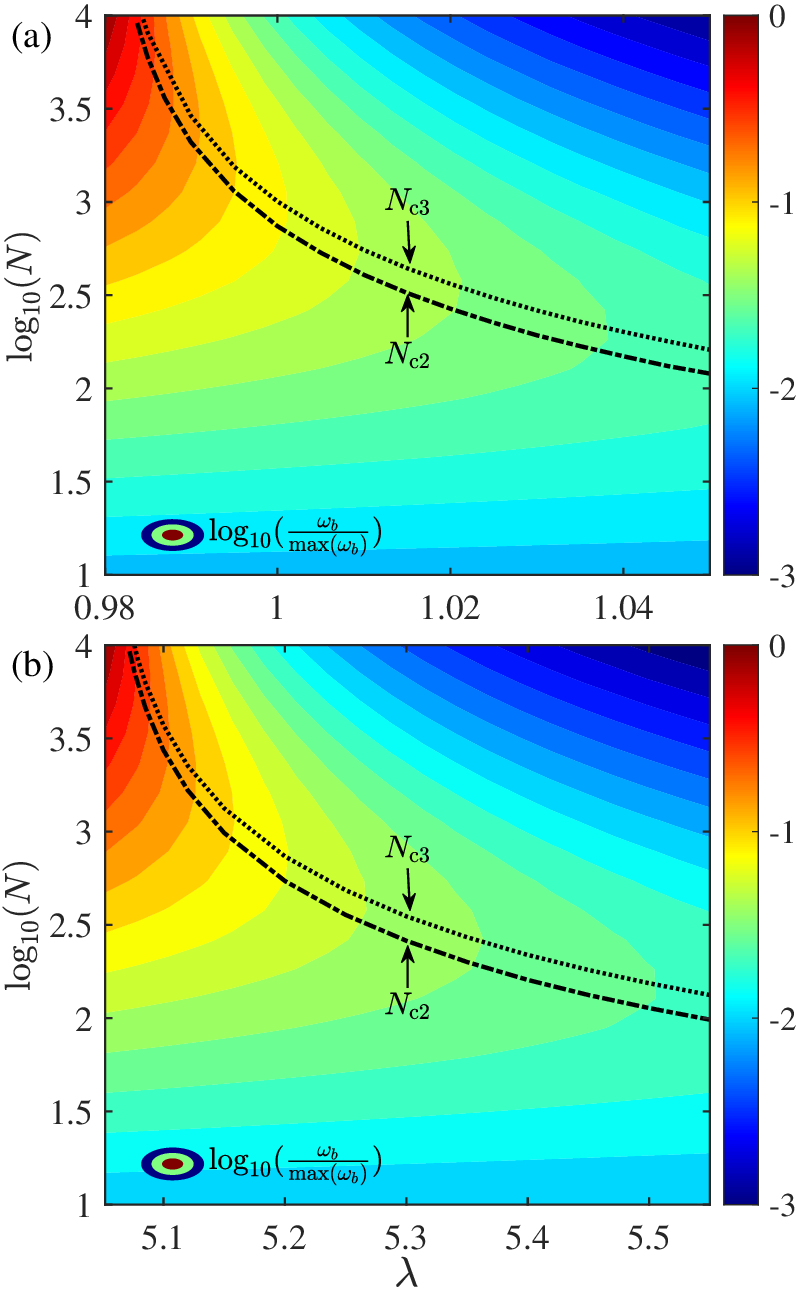}
\caption{\justifying
Breathing-mode frequency in the parameter space of interaction strength ratio $\lambda$ and total atom number $N$. (a), (b) The contour plots of the breathing-mode frequency of quantum droplets in the $\lambda$-$\log_{10}(N)$ plane at the parameter regions I and II, respectively. The gradient color represents the scaled breathing mode frequency $\log_{10}{[\omega_\mathrm{b}/\mathrm{max}(\omega_\mathrm{b})]}$. Here, the dash-dotted and dotted lines, i.e., $N_{c2}$ and $N_{c3}$, mark the minimum of the root-mean-square radius and the maximum of the breathing-mode frequency, respectively. The interaction parameters are the same as in Fig.~\ref{fig:phasediagram}.} 
\label{fig:breathingmode}
\end{figure}

%%%%%%%%%%%%%%%%%%%%%%%%%%%%%%%%%%%%%%%%%%%%%%%%%%%%%%
%%%%%%%%%%%%%%%%%%%%%%%%%%%%%%%%%%%%%%%%%%%%%%%%%%%%%%
\section{RESULTS AND DISCUSSIONS} \label{sec: results}
\begin{figure*}[ht]
\centering
\includegraphics[width=0.96\textwidth]{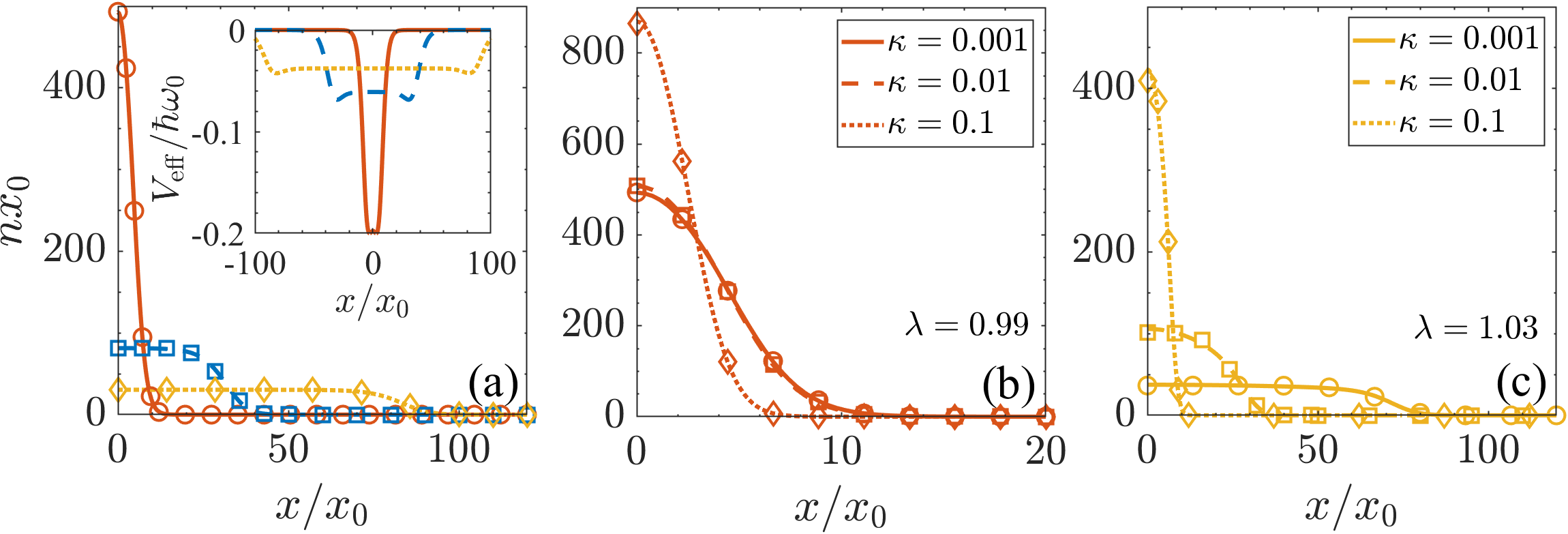}
\caption{\justifying
Total density distribution $n$ at various values of interaction ratio $\lambda=g_1/g_2$ and trapping potential strength $\kappa$. (a) Density profiles without external potential (i.e., $\kappa=0$) at three typical values of $\lambda=0.99$ (red solid line), $1.01$ (blue dashed line), $1.03$ (yellow dotted line) in parameter region I. The corresponding results obtained from the variational approach are shown by the circles, squares, and diamonds. The inset shows the corresponding effective potentials owing to the mean-field and Lee-Huang-Yang energy contributions.
(b), (c) Density profiles for $\lambda=0.99$ and $\lambda=1.03$ at three values of potential strength $\kappa=0.001$, $0.01$, and $0.1$, denoted by the solid, dashed, and dash-dotted curves, respectively. }
\label{fig:densityprofile} 
\end{figure*}

As in the experiments with cold atoms~\cite{gorlitz2001realization, schreck2001quasipure, greiner2001exploring}, we can carefully manipulate the confining frequencies of external harmonic traps to achieve the cigar-shaped system or 1D geometry adopted in this work. Hence, by flexibly tuning the confinement trapping frequencies $(\omega_x,\omega_y=\omega_z=\omega_{\perp})$, we can realize the quasi-1D geometry along the $x$ axis with the effective 1D interaction strength $g^\mathrm{(1D)}=\frac{4\hbar^2a_{s}}{ml_{\perp}^2}(1-\mathcal{C}\frac{a_{s}}{l_{\perp}})^{-1}$~\cite{olshanii1998atomic}, with $\mathcal{C}\approx1.4603$, or $g^\mathrm{(1D)}\approx\frac{\hbar^2a_{s}}{2ml_{\perp}^2}$~\cite{petrov2016ultradilute}, with the oscillator length $l_{\perp}$ in the tightly confined $y$-$z$ plane ($\omega_{\perp}\gg\omega_x$). By taking two groups of quasi-1D confinement closely related to the experiments, we obtain two sets of dimensionless effective 1D interaction strengths $g_2=0.100,g_{12}=-0.099$, $g_{1}\in[0.098,0.105]$ and $g_2=0.089,g_{12}=-0.200$, $g_{1}\in[0.449,0.494]$, giving rise to two groups of the interaction ratio with $\delta g/g\ll1$, i.e., region I with $\lambda\in[0.98,1.05]$ near unity and region II with $\lambda\in[5.05,5.55]$ away from unity.
To further consider a binary bosonic gas with tunable interactions, we can utilize the Feshbach resonance of $^{39}$K atoms at $B_0=58.92$G where the intra- and interspecies $s$-wave scattering lengths $a_{22}$ and $a_{12}$ in three dimensions are approximately constant while the intraspecies scattering length $a_{11}$ spans over a wide range~\cite{derrico2007feshbach}. Taking a typical harmonic confinement $(\omega_x,\omega_{\perp})=2\pi\times[300,1.85\times10^4]$Hz from Ref.~\cite{greiner2001exploring} for a binary $^{39}$K gas, the calculated interaction ratio spans the range $[0.21,7.09]$ with magnetic field $B$ in $[50,58]$G, which covers both parameter sets used in this work.

%%%%%%%%%%%%%%%%%%%%%%%%%%%%%%%%%%%%%%%%%%%%%%%%%%%%%%
\subsection{Phase diagram}

In Fig.~\ref{fig:phasediagram}, we present the typical phase diagrams of the free-space (i.e., $\kappa=0$) quantum droplets in the asymmetrically interacting Bose-Bose mixture modulated by the interaction strength ratio $\lambda\equiv g_1/g_2$ and the total atom number $N$ in two parameter regions I [Fig.~\ref{fig:phasediagram}(a)] and [Fig.~\ref{fig:phasediagram}(b)] II. The scaled peak density $\log_{10}{[n_\mathrm{peak}/\mathrm{max}(n_\mathrm{peak})]}$ in Fig.~\ref{fig:phasediagram} is calculated to illustrate typical phases in gradient colors. Here, we have introduced a critical atom number $N_{c1}$ to distinguish the phase boundary of the typical Gaussian-like and flat-top phases. $N_{c1}$ is determined by the position where the peak density of the quantum droplet becomes saturated, which exhibits a monotonically decreasing behavior with respect to the ratio $\lambda$, shown as the dashed line in Fig.~\ref{fig:phasediagram}.
As $N$ rises, the system manifests as a Gaussian-like quantum phase with a continuously increasing peak density and a varying cloud size. Above $N_{c1}$, the peak density remains unchanged, as shown by the vertically unchanged color above the dashed lines. This reveals the dominant contribution of the MF and LHY energies, and the system steps into a flat-top quantum droplet phase with a constant peak density. 

In the same parameter regions I and II, we further calculate the breathing-mode frequency and illustrate the contour plots of the scaled breathing-mode frequency $\log_{10}{[\omega_\mathrm{b}/\mathrm{max}(\omega_\mathrm{b})]}$ in Fig.~\ref{fig:breathingmode}. Two curves, namely $N_{c2}$ and $N_{c3}$, are introduced and exhibit a monotonically decreasing behavior as the ratio $\lambda$ rises. These two curves, represented by the dash-dotted and dotted lines, are relatively very close, corresponding to the minimum of the root-mean-square radius and the maximum of the breathing-mode frequency, respectively. When $N$ increases gradually towards these two curves, the role of the MF and LHY energies starts to become non-negligible and competes against the quantum pressure induced by the single-particle term. As a result, the system experiences a transition from a Gaussian-like phase to a flat-top phase. Near $N_{c2}$ or $N_{c3}$, the effectively attractive MF and LHY energies compress the droplet against the repulsive quantum pressure to reach the minimum radius in the density profile. Meanwhile, the breathing-mode frequency is pronouncedly enhanced and reaches a maximum; see Fig.~\ref{fig:breathingmode} and details in the later section.

%%%%%%%%%%%%%%%%%%%%%%%%%%%%%%%%%%%%%%%%%%%%%%%%%%%%%%
\subsection{Stationary properties}
\begin{figure}[t]
\centering
\includegraphics[width=0.48\textwidth]{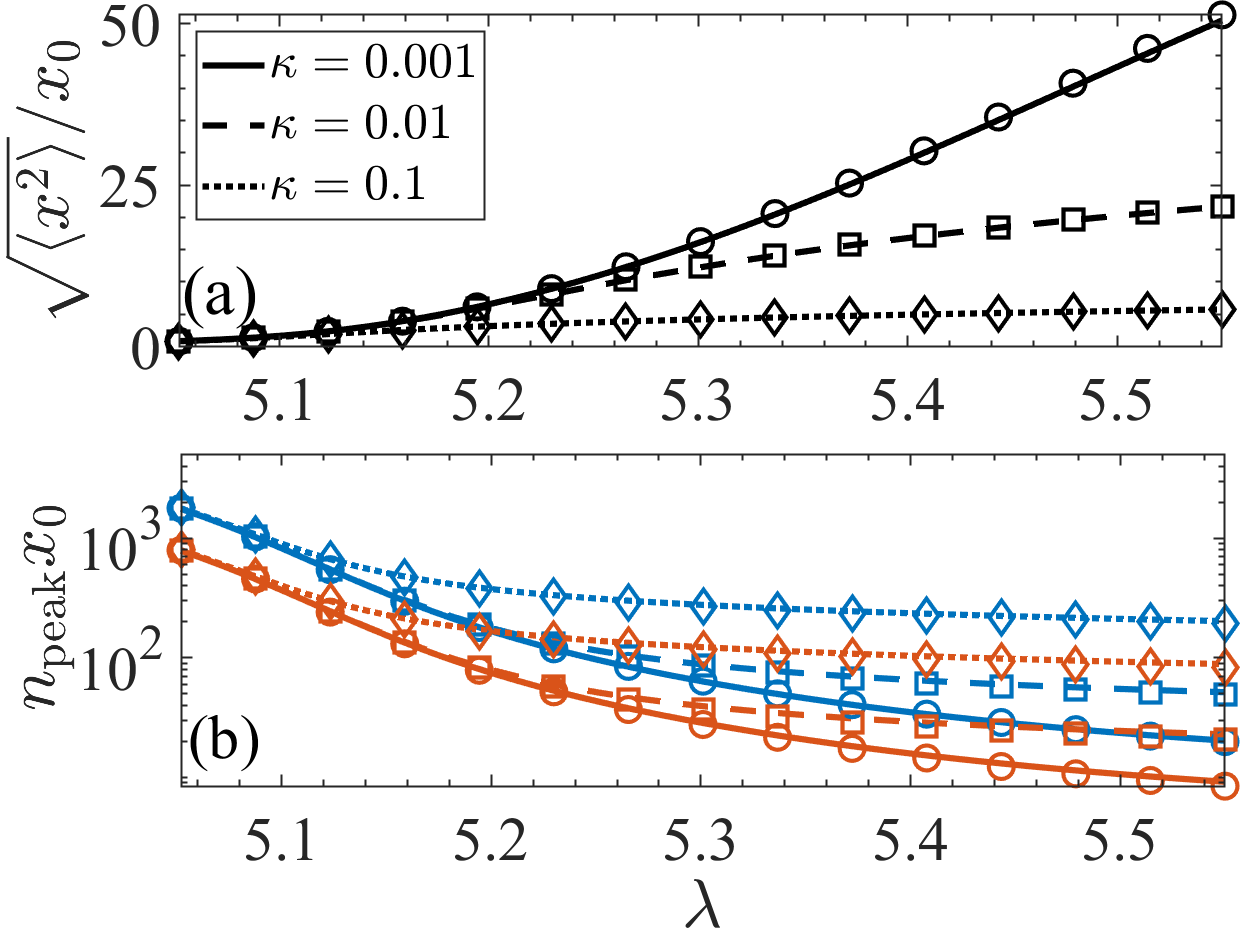}
\caption{\justifying
(a) The root-mean-square radius $\sqrt{\left< x^2 \right>}$ and (b) the peak density $n_\mathrm{peak}$ of two spin components at three values of trapping potential strength $\kappa=0.001$ (solid lines), $0.01$ (dashed lines), and $0.1$ (dotted lines), as functions of the interaction ratio $\lambda$ in parameter region II with $N=5000$. The hollow circles, squares, and diamonds show the corresponding results obtained from the variational approach. The red and blue colors in (b) denote the results for spin components 1 and 2, respectively.}
\label{fig:widthpeak1}
\end{figure}
\begin{figure}[t]
\centering
\includegraphics[width=0.48\textwidth]{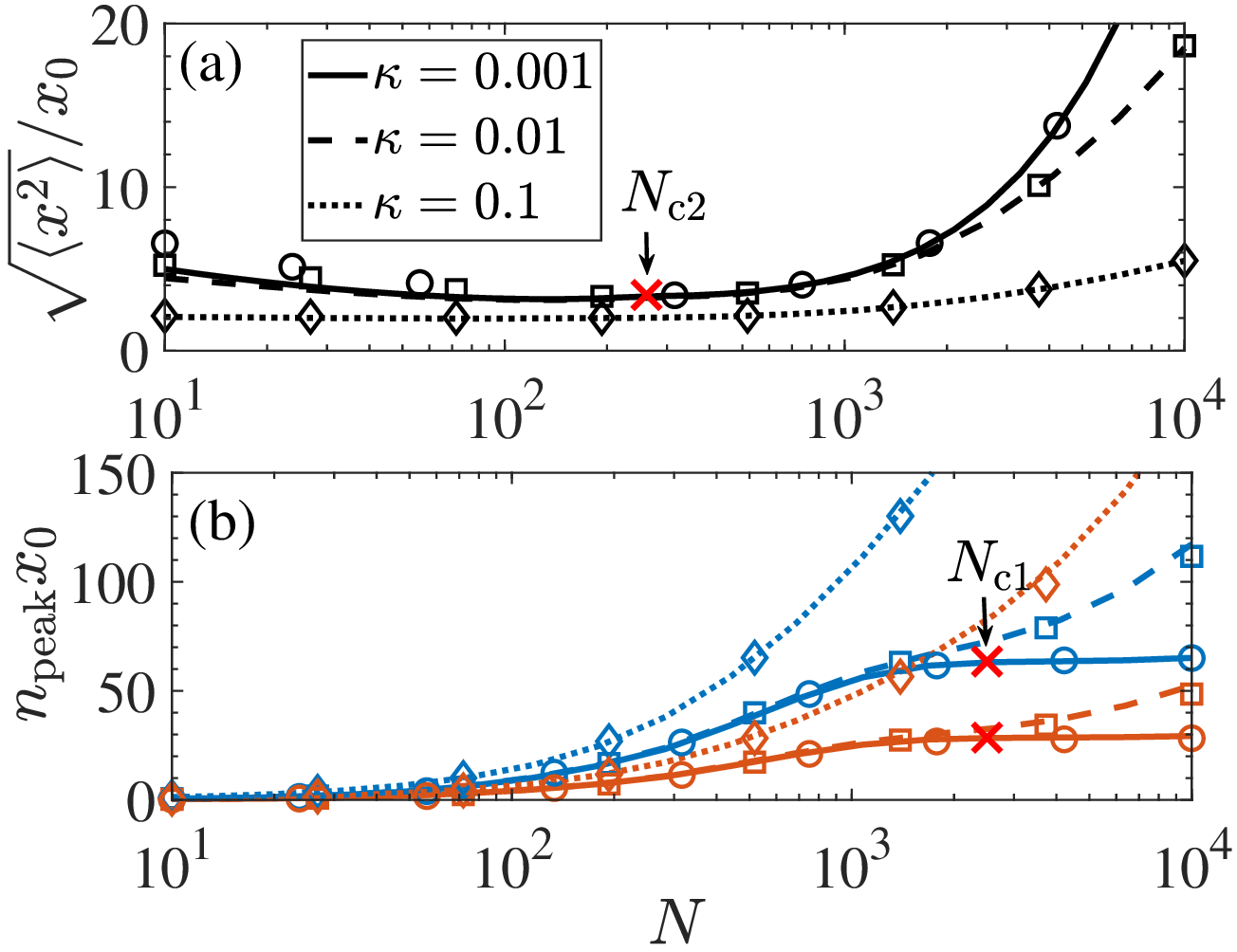}
\caption{\justifying
(a) The root-mean-square radius $\sqrt{\left< x^2 \right>}$ and (b) the peak density $n_\mathrm{peak}$ of two spin components at three values of trapping potential strength $\kappa=0.001$ (solid lines), $0.01$ (dashed lines), and $0.1$ (dotted lines), as functions of the total atom number $N$ with $\lambda=5.3$. The hollow circles, squares, and diamonds show the corresponding results obtained from the variational approach. The red and blue colors in (b) denote the results for spin components 1 and 2, respectively. Here, the critical atom numbers $N_{c1}$ and $N_{c2}$ are determined by the saturated peak density of the quantum droplet, and the minimum of the radius, respectively.} 
\label{fig:widthpeak2}
\end{figure}

In Fig.~\ref{fig:densityprofile}, we present the ground-state density distributions at typical values of interaction ratio $\lambda$ and trapping potential strength $\kappa$ in parameter region I. In accordance with recent experiments on quantum droplets~\cite{cabrera2018quantum,cheiney2018bright,semeghini2018self,derrico2019observation,ferioli2019collisions}, we first focus on the parameter space near the symmetric point (i.e., $\lambda=1$ or $g_1=g_2$) with the total atom number $N=5000$. In free space, as shown in Fig.~\ref{fig:densityprofile}(a), the effective potential $V_\mathrm{eff}$ in Eq.~\eqref{eq:eGPE} contributed by the mean-field interaction and LHY correction changes from a harmonic-potential shape to a box like shape as $\lambda$ rises. As a consequence, the total density of the quantum droplet exhibits a Gaussian-like profile at $\lambda=0.99$ (red solid curves), and turns into a flat-top structure at larger $\lambda$ (blue dashed and yellow dotted curves). This gradual transition in the density of droplets from a Gaussian-like to flat-top structure is reminiscent of the same condition as the total atom number increases in previous works~\cite{astrakharchik2018dynamics,otajonov2019stationary,hu2020collective,tylutki2020collective,parisi2020quantum,mistakidis2021formation,sturmer2021breathing,dong2022internal,du2023ground,fei2024collective}, revealing the characteristic signature of a quantum phase transition in self-bound droplets. These three typical density profiles are located at different positions in the parameter space, as indicated by the hollow circle, square, and diamond in Fig.~\ref{fig:phasediagram}(a).
In Fig.~\ref{fig:densityprofile}(b) and Fig.~\ref{fig:densityprofile}(c), the density profiles are present at three values of trapping potential strength $\kappa=0.001$, $0.01$, and $0.1$, for $\lambda=0.99$ (i.e., red lines) and $\lambda=1.03$ (i.e., yellow lines), respectively. At small $\lambda=0.99$, the Gaussian-like density profile is gradually compressed by the enhancing trapping potential as $\kappa$ increases, maintaining the Gaussian structure with a much more localized shape and a higher peak density, as shown in Fig.~\ref{fig:densityprofile}(b). In contrast, the flat-top density for $\lambda=1.03$ in free space is significantly modified by the increasing $\kappa$, transforming from a flat-top structure to a Gaussian-like distribution with the radius remarkably squeezed owing to the relatively strong trapping potential; see Fig.~\ref{fig:densityprofile}(c). Intriguingly, the analytical solutions denoted by the hollow symbols obtained from the super-Gaussian variational ansatz in Eq.~\eqref{eq:ansatz} show excellent agreement with the numerical results denoted by the curves across all parameter regimes, as shown in Fig.~\ref{fig:densityprofile}.

\begin{figure*}[ht]
\centering
\includegraphics[width=0.96\textwidth]{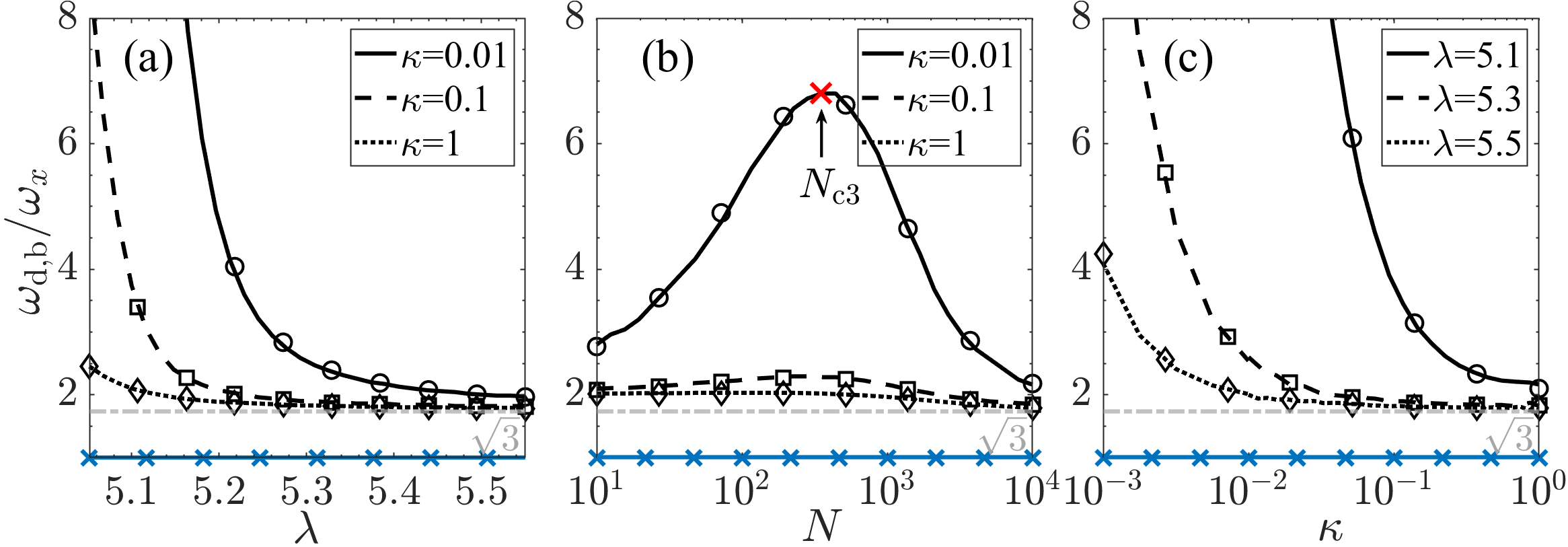}
\caption{\justifying
The dipole-mode (blue lines) and breathing-mode (black lines) frequencies $\omega_\mathrm{d,b}$, as functions of (a) interaction ratio $\lambda$ with $N=5000$, (b) total atom number $N$ with $\lambda=5.3$, and (c) trapping potential strength $\kappa$ with $N=5000$. The solid, dashed, and dotted lines in (a) and (b) represent three sets of trapping potential strength, $\kappa=0.01$, $0.1$, and $1$, respectively, while they denote $\lambda=5.1, 5.3, 5.5$ in (c). The hollow circles, squares, and diamonds show the corresponding results obtained from the variational approach.
Here, the horizontal gray dashed line indicates the analytic breathing mode frequency $\omega_\mathrm{b}=\sqrt{3}\omega_x$ for a conventional weakly interacting Bose gas. The critical atom number $N_{c3}$ in the middle plot is determined by the maximum of the breathing-mode frequency.}
\label{fig:dipolebreathing}
\end{figure*}
To characterize the typical phases in quantum droplets, we turn to investigate the root-mean-square radius $\sqrt{\langle x^2 \rangle}$ and the peak density $n_\mathrm{peak}$ of two spin components as functions of the interaction ratio $\lambda$ and total atom number $N$. In Fig.~\ref{fig:widthpeak1}, the radius and peak density calculated by solving the extended GPE are presented as functions of $\lambda$ in parameter region II at three values of trapping potential strength $\kappa=0.001$, $0.01$, and $0.1$, as shown by the solid, dashed, and dotted curves, respectively. In general, the radii of two spin components [i.e., black curves in Fig.~\ref{fig:widthpeak1}(a)] are the same while their peak densities [i.e., red and blue curves in Fig.~\ref{fig:widthpeak1}(b)] are not since the value of $\lambda$ is far away from unity, giving rise to different atom numbers of two spin components. In nearly free space, i.e., at sufficiently weak trapping potential with $\kappa=0.001$, the radius in the black solid curve increases monotonically with $\lambda$ as the system transitions from a Gaussian-like phase to a flat-top phase, consistent with the scenario illustrated in Fig.~\ref{fig:densityprofile}(a). As a result of a fixed total atom number, the two peak densities decrease gradually and tend to saturate at a specific value. This is because at relatively larger $\lambda$, the system exhibits a flat-top structure in the density profile with a nearly constant peak density and an expanding radius, as described in previous works~\cite{astrakharchik2018dynamics, tylutki2020collective,parisi2020quantum}. At relatively stronger trapping potentials, i.e., $\kappa=0.01$ and $0.1$, shown by dashed and dotted curves respectively, the variations in the radius and peak density become much more gentle and smooth. This can be understood as a consequence of the stronger confinement imposed by the external potential, which compresses the droplet into a smaller width and thus elevates its central density. The corresponding results obtained from the variational approach are denoted by the hollow symbols in Fig.~\ref{fig:widthpeak1}, which intriguingly show a quantitative agreement with the numerical ones.

In Fig.~\ref{fig:widthpeak2}, we further show the dependence of the radius and the peak density of two spin components on total atom number $N$ at different strengths of external trapping potential. In nearly free space with $\kappa=0.001$, the radius exhibits a nonmonotonic behavior with respect to $N$, i.e., decreasing gradually to a minimum and then turning to increase significantly, as shown by the black solid line in Fig.~\ref{fig:widthpeak2}(a). Meanwhile, the peak densities of both spin components increase monotonically and turn to saturate at the specific values as $N$ rises, as shown by the red and blue solid lines in Fig.~\ref{fig:widthpeak2}(b). These behaviors are reminiscent of our previous findings in Ref.~\cite{du2023ground}. At small $N$, the quantum pressure arising from the dominant kinetic energy pushes the atoms away to maintain a relatively large width. As $N$ increases, a competing effectively attractive interaction owing to the MF and LHY energy terms will slightly squeeze the condensate, and hence the radius relatively decreases. We introduce the first critical value $N_{c2}$ in Fig.~\ref{fig:widthpeak2}(a), where the radius reaches the minimum; see, also, the black dash-dotted lines in Fig.~\ref{fig:phasediagram}. At relatively large $N$, the system dominated by the LHY energy steps into the flat-top droplet phase characterized by a saturated peak density and an increasing width depending linearly on $N$. Similarly, the critical $N_{c1}$ in Fig.~\ref{fig:widthpeak2}(b) indicates the position where the peak density becomes constant, which is denoted by the black dashed lines in Fig.~\ref{fig:phasediagram}. At relatively stronger trapping potentials, i.e., $\kappa=0.01$ and $0.1$ shown by dashed and dotted curves respectively, the condensates are significantly compressed by the external confinement. As a result, the radius tends to be flat with the non-monotonic behavior disappearing, and the peak densities rise notably showing a linear dependence on $N$. Our results from the variational approach, presented in the hollow symbols in Fig.~\ref{fig:widthpeak2}, show a good agreement with the numerical ones.

%%%%%%%%%%%%%%%%%%%%%%%%%%%%%%%%%%%%%%%%%%%%%%%%%%%%%%
\subsection{Collective excitation modes}

Now we turn to the collective elementary excitations or collective modes of quantum droplets in this asymmetrically interacting 1D Bose-Bose mixture, which reveal the important dynamical properties of this novel phase and can be well probed in ultracold-atomic experiments.
\begin{figure}[t]
\centering
\includegraphics[width=0.48\textwidth]{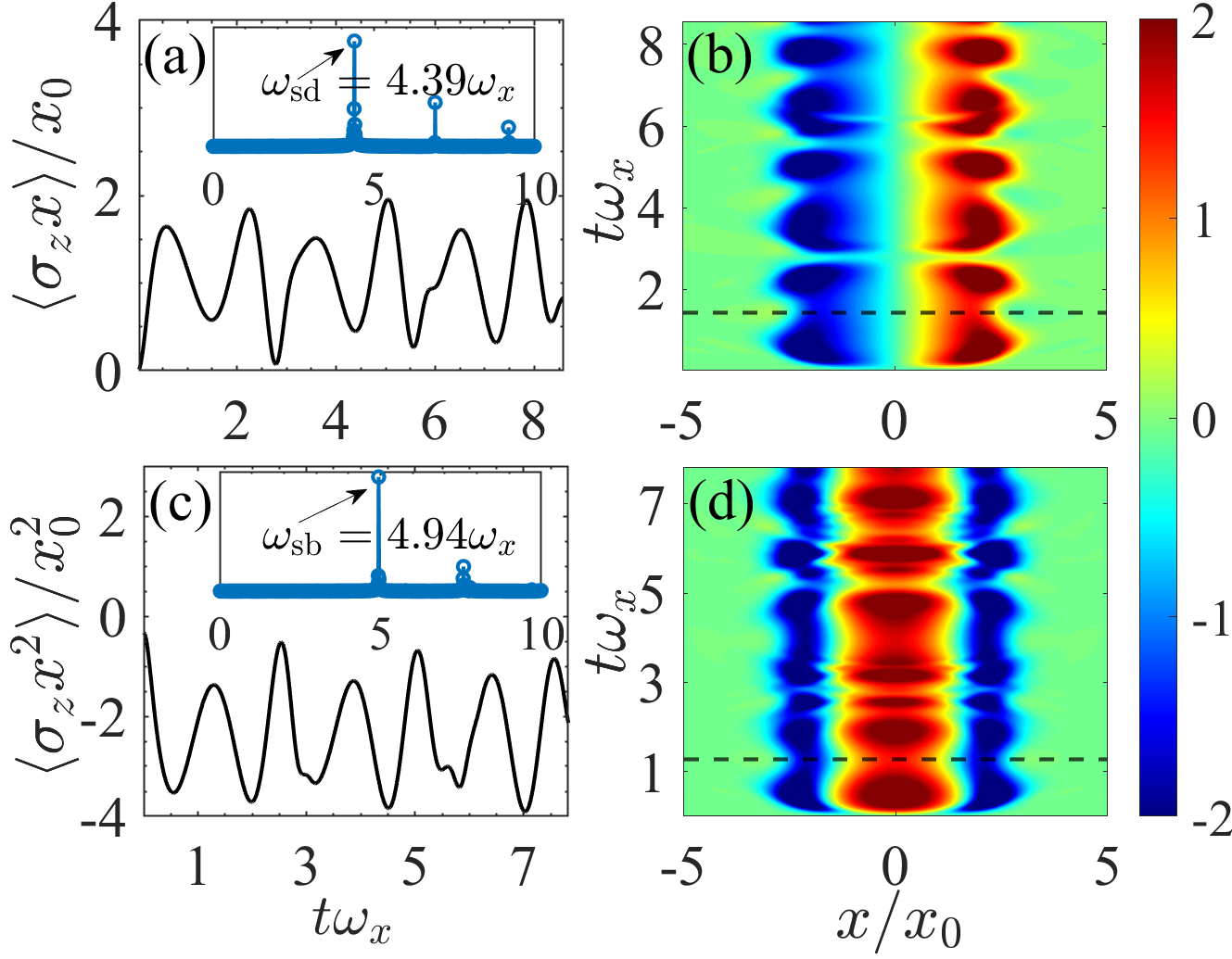}
\caption{\justifying
Illustration of typical time-dependent observables $\left<\sigma_z x \right>, \left<\sigma_z x^2 \right>$ and contour plots of the spin-density distribution $\delta n(x,t)=n_{1}-n_{2}$ for (a), (b) spin-dipole mode and (c), (d) spin-breathing mode. The Fourier analyses of the observables are shown in the insets, while the dominant peaks at $\omega_\mathrm{sd}=4.39\omega_x$ and $\omega_\mathrm{sb}=4.94\omega_x$ extract the spin-dipole- and spin-breathing-mode frequencies, respectively. The corresponding time durations of one period, i.e., $2\pi/\omega_\mathrm{sd}\approx1.43/\omega_x$ and $2\pi/\omega_\mathrm{sb}\approx1.27/\omega_x$, are denoted by the black dashed lines in (b),and (d).}
\label{fig:relativedynamics}
\end{figure}
\begin{figure}[t]
\centering
\includegraphics[width=0.48\textwidth]{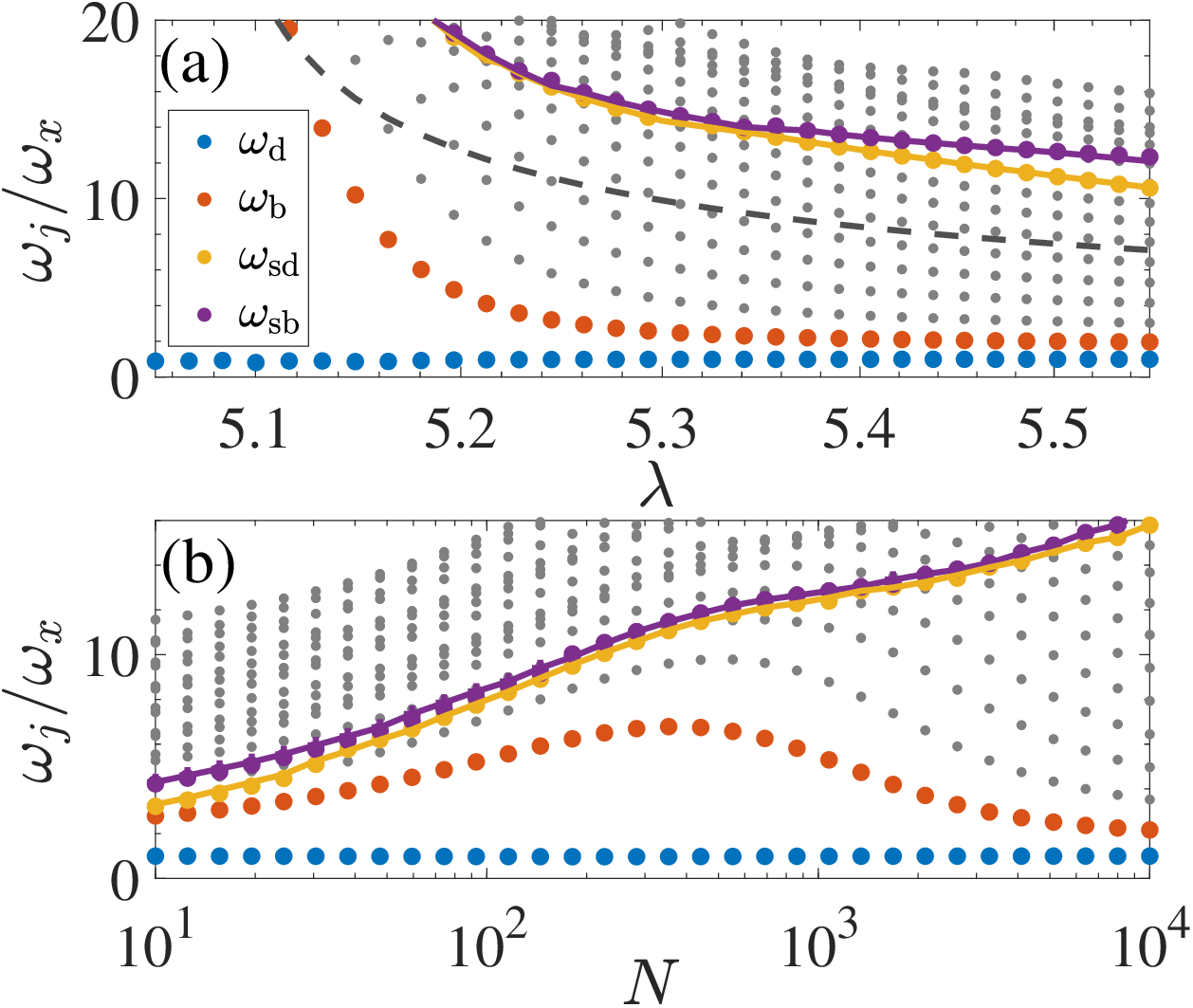}
\caption{\justifying
Collective excitation frequencies $\omega_j$ for $\kappa=0.01$ in parameter region II, as a function of (a) interaction ratio $\lambda$ and (b) total atom number $N$ with $\lambda=5.3$, calculated by the linearization technique in Eq.~\eqref{eq:Linearization} (i.e., dot symbols). Here, the frequencies of the dipole, breathing, spin-dipole, and spin-breathing modes are highlighted by blue, red, yellow, and purple dots, respectively. Two spin modes show an excellent agreement with the one obtained from the time-dependent extended GPE, denoted by the solid lines in yellow and purple.
The gray dashed lines indicate the sum-rule prediction of the spin-dipole-mode frequency, i.e., Eq.(3) in Ref.~\cite{geier2021exciting}, for the conventional weakly interacting two-component Bose gas. Other parameters are identical to those in Fig.~\ref{fig:dipolebreathing}.}
\label{fig:spinmodes}
\end{figure}

In Fig.~\ref{fig:dipolebreathing}, the frequencies of the dipole and breathing modes are shown as functions of the interaction ratio $\lambda$ in parameter region II, the total atom number $N$, and the trapping potential strength $\kappa$. Here, the curves indicate the numerical calculations via the extended GPE, while the hollow symbols denote the corresponding results obtained from the sum-rule approach with a variational approximation. In Fig.~\ref{fig:dipolebreathing}(a), the breathing-mode frequency $\omega_\mathrm{b}$ in nearly free space (i.e., black solid lines) will significantly increase as $\lambda$ decreases and tends to diverge at the left end. This is because at the left end of $\lambda$, the $\delta g$ tends to be zero, and thus the repulsive MF energy disappears and the attractive LHY energy leads the system to collapse. Moreover, larger trapping strengths $\kappa$, i.e., black dashed and dotted lines, lead to a more gradual decline in $\omega_\mathrm{b}$, which asymptotically approaches the result of a weakly interacting 1D Bose gas $\omega_\mathrm{b} = \sqrt{3}\omega_x$~\cite{ dalfovo1999theory,menotti2002collective} (horizontal gray dash-dotted line) at the right side of $\lambda$. 
This can be understood as follows. As the ratio $\lambda$ increases, the system approaches the limit of a single-component BEC dominated by the intraspin interaction $g_1$, while $g_2$ and $g_{12}$ remain relatively weak and fixed. In this regime, the system moves away from the collapse condition $\delta g \sim 0$ (which corresponds to relatively small $\lambda$) and the LHY correction becomes less significant. Consequently, the system behaves more like a conventional weakly interacting Bose gas confined in a harmonic trap.
In Fig.~\ref{fig:dipolebreathing}(b), the breathing-mode frequency of droplets in nearly free space (i.e., black solid lines) shows a pronounced non-monotonic behavior as the atom number $N$ rises, i.e., significantly increasing, reaching the maximum and turning to decreasing sharply towards the conventional 1D Bose gas limit, reproducing the results in previous works~\cite{du2023ground,menotti2002collective,kimura2002Breathing}. Similarly, a stronger trapping potential plays a more important role and smooths out the hump in the frequency, as shown by the black dashed and dotted lines. Here, we also introduce a critical atom number $N_{c3}$ in Fig.~\ref{fig:dipolebreathing}(b) where the MF and LHY energies play a competing role against the quantum pressure and the breathing mode frequency reaches the maximum; see, also, the black dotted lines in Fig.~\ref{fig:phasediagram}. In Fig.~\ref{fig:dipolebreathing}(c), the mode frequencies are illustrated as a function of trapping strength $\kappa$ for three values of interaction ratio $\lambda$. As the trapping potential gets stronger, the breathing-mode frequency significantly decreases towards the conventional Bose gas limit. In addition, a smaller $\lambda$ leads $\delta g$ closer to zero, and thus the system is more favorable to collapse and $\omega_\mathrm{b}$ becomes divergent more quickly as $\kappa$ decreases. 
Here, for relatively larger total atom numbers $N$ or stronger trapping potentials $\kappa$ as shown in Fig.~\ref{fig:dipolebreathing}(b) and Fig.~\ref{fig:dipolebreathing}(c), the interaction and external trapping energies dominate over the quantum pressure arising from the kinetic energy associated with spatial variations of the cloud. The droplet profile becomes significantly compressed and the system enters the Thomas–Fermi regime, characterized by a BEC-like density distribution (see, e.g., the density profiles in our previous works in Refs.~\cite{du2023ground,fei2024collective}). In this regime, the system's behavior is governed primarily by the external confinement rather than by the delicate balance between mean-field and LHY corrections. As a result, the breathing-mode frequency tends to approach that of a harmonically trapped BEC.
In all figures, the dipole-mode frequency $\omega_\mathrm{d}$ equals the harmonic trapping frequency over all the ranges guaranteed by the Kohn theorem, i.e., $\omega_\mathrm{d}=\omega_x$, as shown by the numerical calculations (blue lines) and the analytic results (blue crosses). Here, our numerical calculations agree well with the analytic results in Eq.~\eqref{eq:omega_d} and Eq.~\eqref{eq:omega_b}, denoted by the hollow symbols.

In addition to the above modes governed by the motion of the total density, we further study the spin-dependent collective excitations, i.e., spin-dipole and spin-breathing modes, by introducing spin-dependent perturbations in Eq.~\eqref{eq:pertubation} and analyzing the related time-dependent observables. By perturbing the system with a small spin-dependent displacement $\sigma_z\delta x$, the spin-dipole mode is excited and can be studied through the time evolution of the associated spin center-of-mass displacement $\left<\sigma_zx(t)\right>$, as shown in Fig.~\ref{fig:relativedynamics}(a). The inset shows its Fourier analysis $\tilde{f}(\omega)$, with some peaks indicating the existence of multiple modes in the related dynamics. Here, the highest peak corresponds to the spin-dipole mode at $\omega_\mathrm{sd}=4.39\omega_x$, corresponding to an oscillation period of $T_\mathrm{sd} = 2\pi / \omega_\mathrm{sd} \approx 1.43/\omega_x$. This can be further verified by the spatiotemporal evolution of the spin-density distribution $\delta n(x,t) = n_{\uparrow}(x,t) - n_{\downarrow}(x,t)$ in Fig.~\ref{fig:relativedynamics}(b). The relative motion between the two spin components leads to alternating high- and low-density differences at opposite ends of the cloud, and the higher-frequency modes give rise to small revisions in spin density. As a result, the spin-density distribution exhibits a characteristic antisymmetric profile that oscillates periodically in time whose period agrees well with $T_\mathrm{sd}$ shown as a black dashed line. Similarly, we excite the spin breathing mode by introducing a spin-dependent variation of the trapping frequency and show the evolution of the related spin squared radius $\left<\sigma_zx^2\right>$ in Fig.~\ref{fig:relativedynamics}(c). The inset reveals that the spin breathing mode takes a frequency of $\omega_\mathrm{sb}=4.94\omega_x$. This period $T_\mathrm{sb} = 2\pi / \omega_\mathrm{sb} \approx 1.27/\omega_x$ is further confirmed in the spatiotemporal spin density as shown by the black dashed line in Fig.~\ref{fig:relativedynamics}(d). The density of different spin components exhibits distinct alternating behaviors at the center and the tails, i.e., relative trends of expansion and contraction, with the difference in their relative motion reaching a maximum at the half cycle and returning to a minimum at the full cycle. As a result, the middle and both end regions of the spin-density distribution display structures with opposite signs that are symmetric about the origin $x=0$.

Based on the procedure described in the last paragraph and Fig.~\ref{fig:relativedynamics}, the frequencies of the spin-dipole mode (i.e., solid yellow lines) and spin-breathing mode (i.e., solid purple lines) are extracted and depicted in Fig.~\ref{fig:spinmodes}, as functions of the interaction ratio $\lambda$ and total atom number $N$ at a trapping potential strength $\kappa=0.01$. 
Meanwhile, we have applied the linearization technique presented in Sec.~\ref{sec:linearization} to calculate the low-energy collective excitation spectrum denoted by the dot symbols, where the concerned dipole, breathing, spin-dipole, and spin-breathing modes are highlighted in blue, red, yellow, and purple dots, respectively. In practice, we identify the specific density and spin modes within the excitation spectrum (i.e., the dots in Fig.~\ref{fig:spinmodes}) by examining the real-time evolution of the quasiparticle densities $|\eta_j|^2(x,t) = |\eta_{j,1}|^2 + |\eta_{j,2}|^2$ and $\delta |\eta_j|^2(x,t) = |\eta_{j,1}|^2 - |\eta_{j,2}|^2$ for the $j$th mode with frequency $\omega_j$, and classifying its characteristic pattern in the $x$-$t$ contour plot as in Fig.~\ref{fig:relativedynamics}. 
In general, we find that these results of the four mode frequencies show an excellent agreement with the one from numerically solving the time-dependent extended GPE.
In Fig.~\ref{fig:spinmodes}(a), both spin-mode frequencies show a monotonic dependence on $\lambda$, i.e., decreasing sharply as $\lambda$ rises and tending to saturate. This trend of saturation becomes more pronounced for larger $\kappa$. This is because the relatively strong trapping potential will compress the droplet to maintain a Gaussian-like wave function or a Thomas-Fermi density distribution; see, also, the radius in Fig.~\ref{fig:widthpeak1}(a) and the breathing-mode frequency in Fig.~\ref{fig:dipolebreathing}(a). Thus, the spin-dipole-mode frequency tends to approach the sum-rule prediction, i.e., Eq.(3) in Ref.~\cite{geier2021exciting}, of the conventional weakly interacting two-component Bose gas. In Fig.~\ref{fig:spinmodes}(b), the frequencies of both spin modes increase monotonically with $N$, in contrast to the nonmonotonic behavior of the breathing mode in Fig.~\ref{fig:dipolebreathing}(b). A stronger confinement will gradually slow down the upward trend in the frequencies of spin modes.
Note that a recent work~\cite{ritu2026spin} has also investigated the spin excitations of one-dimensional self-bound Bose-Bose droplets presented in the current work.

%%%%%%%%%%%%%%%%%%%%%%%%%%%%%%%%%%%%%%%%%%%%%%%%%%%%%%
%%%%%%%%%%%%%%%%%%%%%%%%%%%%%%%%%%%%%%%%%%%%%%%%%%%%%%
\section{CONCLUSIONS AND OUTLOOKS} \label{sec: conclusions}

In this work, we have systematically investigated the ground-state properties and collective excitations of one-dimensional quantum droplets in weakly interacting Bose-Bose mixtures with asymmetric intraspin interactions. By numerically solving the extended Gross-Pitaevskii equation and employing complementary theoretical methods---namely, a variational approximation, the sum-rule approach, and the linearization technique---we have provided a detailed analysis of both static and dynamic behaviors across a wide range of parameters.

Our results reveal that the asymmetry in intraspin interactions, quantified by the ratio $\lambda \equiv g_1/g_2$, plays a crucial role in shaping the density profiles, root-mean-square radii, peak densities, and excitation spectra of quantum droplets. Similar to the effect of the atom number in previous works, we probe a clear transition from Gaussian-like to flat-top density distributions modulated by increasing $\lambda$, a signature of the quantum droplet phase. This transition is further modulated by the strength of the external confinement $\kappa$ and the total atom number $N$. We find that for a fixed trapping strength $\kappa$, the radii of both components are identical and increase monotonically with the interaction ratio $\lambda$, while their peak densities differ and decrease monotonically with this ratio. In addition, the identical radii of both components in free space exhibit a nonmonotonic behavior with the atom number $N$ while their peak densities increase monotonically and turn to saturate at a constant value. We further introduce three critical points $N_{c1}$, $N_{c2}$, and $N_{c3}$ to distinguish different regimes and phases in the phase diagrams of the system.

Furthermore, we have studied four typical low-energy collective excitation modes in this system, i.e., the well-studied dipole and breathing modes, and the less-studied spin-dipole and spin-breathing modes. In our numerical calculations, specific perturbations are designed to excite certain modes and the corresponding frequencies are extracted from the Fourier analysis of the associated time-dependent observables. The frequencies for the dipole and breathing modes are further verified by the sum-rule analysis based on a super-Gaussian variational ansatz. We find that the dipole mode conforms to Kohn's theorem, oscillating at the trap frequency regardless of interaction asymmetry. In contrast, the breathing mode frequency exhibits a pronounced nonmonotonic dependence on $N$ as well as $\lambda$, reflecting the competition between mean-field, Lee-Huang-Yang, and quantum pressure contributions. The spin-dependent modes, excited via spin-dependent perturbations, reveal rich dynamics in the relative motion of the two components. 
Both spin-dipole and spin-breathing mode frequencies exhibit monotonic dependence on the intraspecies interaction asymmetry ratio and total atom number, and the frequencies tend to approach analytic predictions in the relatively stronger confinement limit. 
These two spin-mode frequencies obtained from numerical simulations of the time-dependent extended GPE are in excellent agreement with those derived from the linearization technique. 

In summary, this work provides a robust theoretical framework for understanding and characterizing asymmetrically interacting quantum droplets, paving the way for further experimental and theoretical advances in the study of beyond-mean-field quantum phenomena. The excellent agreement between numerical extended GPE results and those from the variational, the sum-rule, and the linearization methods underscores the reliability of our methods and the physical insight they provide. 

Further questions, such as extension to higher dimensions, the finite-temperature effect, systems with external effects like spin-orbit coupling, and other physical properties such as phase coherence and superfluidity, remain to be investigated to better understand the exotic self-bound quantum-droplet phase. Moreover, the role of beyond-LHY corrections and the dynamics of droplet formation and stability under time-dependent interactions represent promising avenues for future research.

\begin{acknowledgments}
We acknowledge fruitful discussions with Zengqiang Yu and Jia Wang. This work is supported by the Natural Science Foundation of China (Grants No. 12474492, No. 12461160324, No. 12204413 and No. 12247101), the Science Challenge Project (Grant No. TZ2025017), the Science Foundation of Zhejiang Sci-Tech University (Grant No. 21062339-Y), the Fundamental Research Funds for the Central Universities (Grant No. lzujbky-2025-jdzx07), the Natural Science Foundation of Gansu Province (No. 25JRRA799), and the ‘111 Center’ under Grant No. B20063.
\end{acknowledgments}

\emph{Data availability}---The data that support the findings of this article are openly available~\cite{xiao2026data}.

%\clearpage
\appendix
\section{more details of the variational approach} \label{app:VA-SuperG}

In this appendix, we present more details of the variational approach employed in the main text. The behaviors of the associated variables, as well as the spin atom number ratio, are shown as functions of the interaction ratio.
\begin{figure}[ht]\centering
\includegraphics[width=0.48\textwidth]  {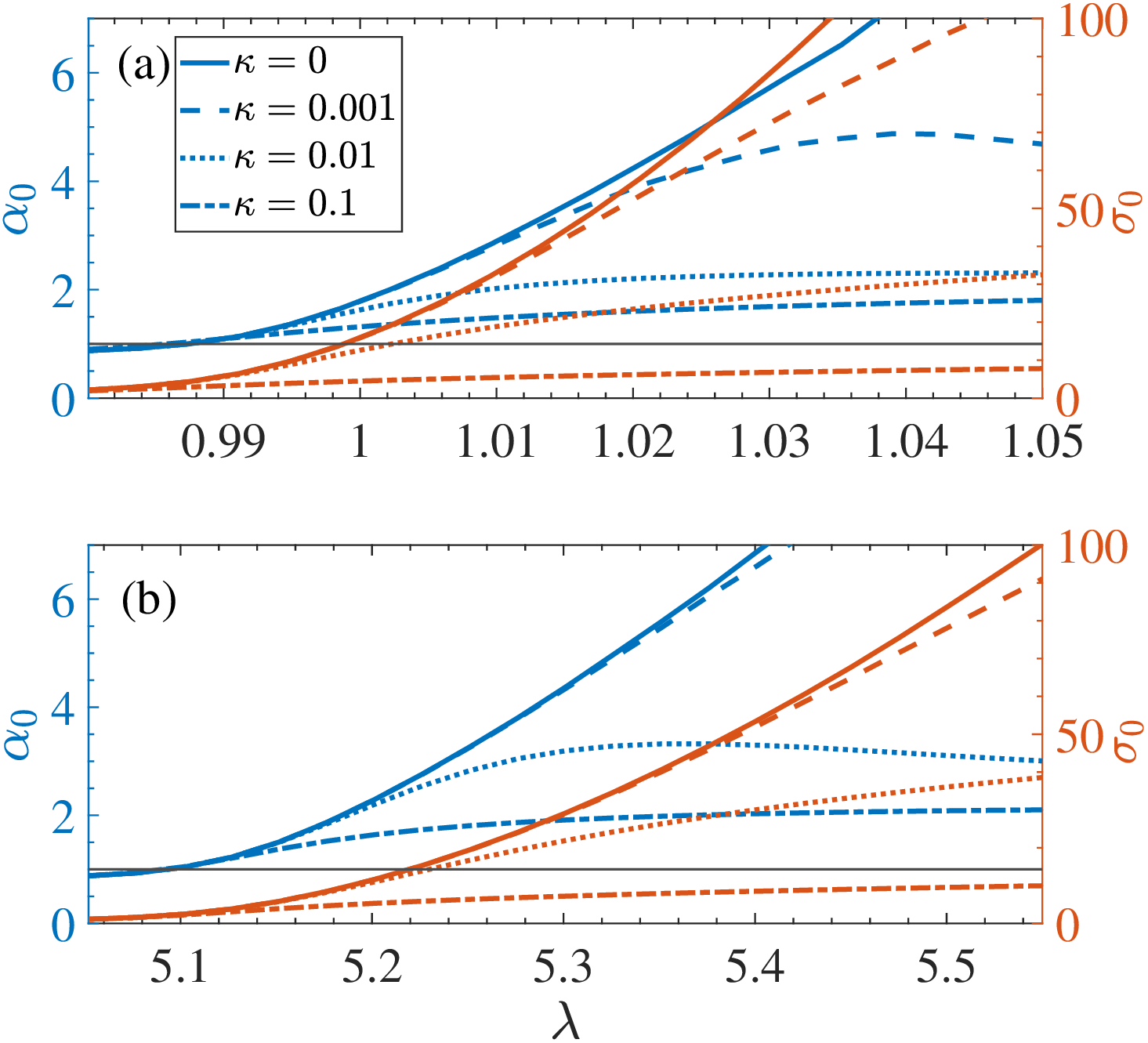}
\caption{\justifying
The extreme points $(\alpha_0,\sigma_0)$ in Eq.~\eqref{eq:ansatz} at the ground states in parameter regions (a) I and (b) II, respectively, as functions of the interaction strength ratio $\lambda$ at different values of trapping potential strength $\kappa = 0$ (solid lines), 0.001 (dashed lines), 0.01 (dotted lines), and 0.1 (dash-dotted lines). The red and blue curves correspond to $\sigma_0$ and $\alpha_0$, respectively.}
\label{figS:variables}
\end{figure}

In our variational approach, we adopt the super-Gaussian variational \emph{ansatz} in Eq.~\eqref{eq:ansatz} for the condensate and substitute it into the energy functional. By minimizing the energy with respect to the variables, i.e., $\frac{\partial{\epsilon_{tot}}}{\partial \alpha}$ and $\frac{\partial{\epsilon_{tot}}}{\partial \sigma}$, we obtain the coupled equations for the extremum points $(\alpha_0,\sigma_0)$ as
\begin{subequations}
    \begin{eqnarray}
    &&-\frac{\hbar^2\alpha_{0}^2\Gamma(2-\frac{1}{2\alpha_{0}})}{m\sigma_{0}^3\Gamma(\frac{1}{2\alpha_{0}})}
    +\frac{m\omega_x^2 \sigma_{0}\Gamma(\frac{3}{2\alpha_{0}})}{\Gamma(\frac{1}{2\alpha_{0}})} \nonumber
  \\
  &&-(\frac{1}{2})^{\frac{1}{2\alpha_{0}}}\frac{N\alpha_{0} \sqrt{\lambda}(g_{12}+g_2\sqrt{\lambda})}{\Gamma(\frac{1}{2\alpha_{0}})\sigma_{0}^2(1+\sqrt{\lambda})^2}\nonumber\\
  && +
  \frac{1}{3}(\frac{2}{3})^{\frac{1}{2\alpha_{0}}}\frac{\sqrt{m}}{\pi\hbar}\sqrt{\frac{N\alpha_{1,0}}{\Gamma(\frac{1}{2\alpha_{0}})\sigma_{0}^3}}(g_2\sqrt{\lambda})^{\frac{3}{2}} = 0,
\end{eqnarray}
\begin{eqnarray}
&&\frac{\hbar^2\Gamma(2-\frac{1}{2\alpha_{0}})}{4m\sigma_{0}^2\Gamma(\frac{1}{2\alpha_{0}})}\left[4\alpha_{0}+\psi^{(0)}(2-\frac{1}{2\alpha_{0}})+\psi^{(0)}(\frac{1}{2\alpha_{0}})\right] 
    \\
    && +
    \frac{m\omega_x^2 \sigma_{0}^2\Gamma(\frac{3}{2\alpha_{0}})}{4\alpha_{0}^2\Gamma(\frac{1}{2\alpha_{0}})}\left[\psi^{(0)}(\frac{1}{2\alpha_{0}})-3\psi^{(0)}(\frac{3}{2\alpha_{0}})\right]\nonumber\\
 && +
    (\frac{1}{2})^{\frac{1}{2\alpha_{0}}}\frac{N\sqrt{\lambda}(g_{12}+g_2\sqrt{\lambda})}{\sigma_{0}(\sqrt{\lambda}+1)^2}\frac{\ln{2}+2\alpha_{0}+\psi^{(0)}(\frac{1}{2\alpha_{0}})}{2\alpha_{0}\Gamma(\frac{1}{2\alpha_{0}})}\nonumber\\
  &&-
  \frac{2}{3}(\frac{2}{3})^{\frac{1}{2\alpha_{0}}}\sqrt{\frac{Nm\alpha_{0}}{\sigma_{0}\Gamma(\frac{1}{2\alpha_{0}})}}\frac{2\alpha_{0}-2\ln{\frac{2}{3}}+\psi^{(0)}(\frac{1}{2\alpha_{0}})}{4\alpha_{0}^2\pi\hbar}(g_2\sqrt{\lambda})^{\frac{3}{2}} = 0, \nonumber
\end{eqnarray}
\end{subequations}
with the zeroth-order Polygamma Function $\psi^{(0)}\left(x\right)\equiv \frac{\Gamma'\left(x\right)}{\Gamma\left(x\right)}$ denoting the differentiation of the Gamma Function $\Gamma(x)$. 
Therefore, after numerically calculating the extremum points $(\alpha_0,\sigma_0)$ from the above coupled equations, we can reconstruct the wave function of the condensate by substituting them back into Eq.~\eqref{eq:ansatz}. In Fig.~\ref{figS:variables}, the variational width $\sigma_0$ and the exponent factor $\alpha _0$ in the variational ansatz are presented as functions of the interaction ratio $\lambda$. 
The horizontal gray solid line indicates the standard Gaussian function corresponding to $\alpha_0=1$. It is clearly seen that the super-Gaussian parameter $\alpha_0$ of the quantum droplet in blue curves significantly increases from around 1 with $\lambda\equiv g_1/g_2$, reflecting a gradual transition of the density profile from a Gaussian-like to flat-top structure. 
At relatively large interaction ratios and at different trapping potential strengths, we can clearly see that the exponent factors $\alpha_0$ in blue curves stay far away from 1. 
Particularly, the exponent factor becomes relatively large in the flat-top droplet regime, where a Gaussian state fails to capture the structural features of flat-top states. This reveals the rationality of the use of the adopted super-Gaussian ansatz with a variable exponent factor here across different parameter regimes instead of a conventional Gaussian function for the condensate wave function.

\begin{figure}[ht]
\centering
\includegraphics[width=0.48\textwidth]  {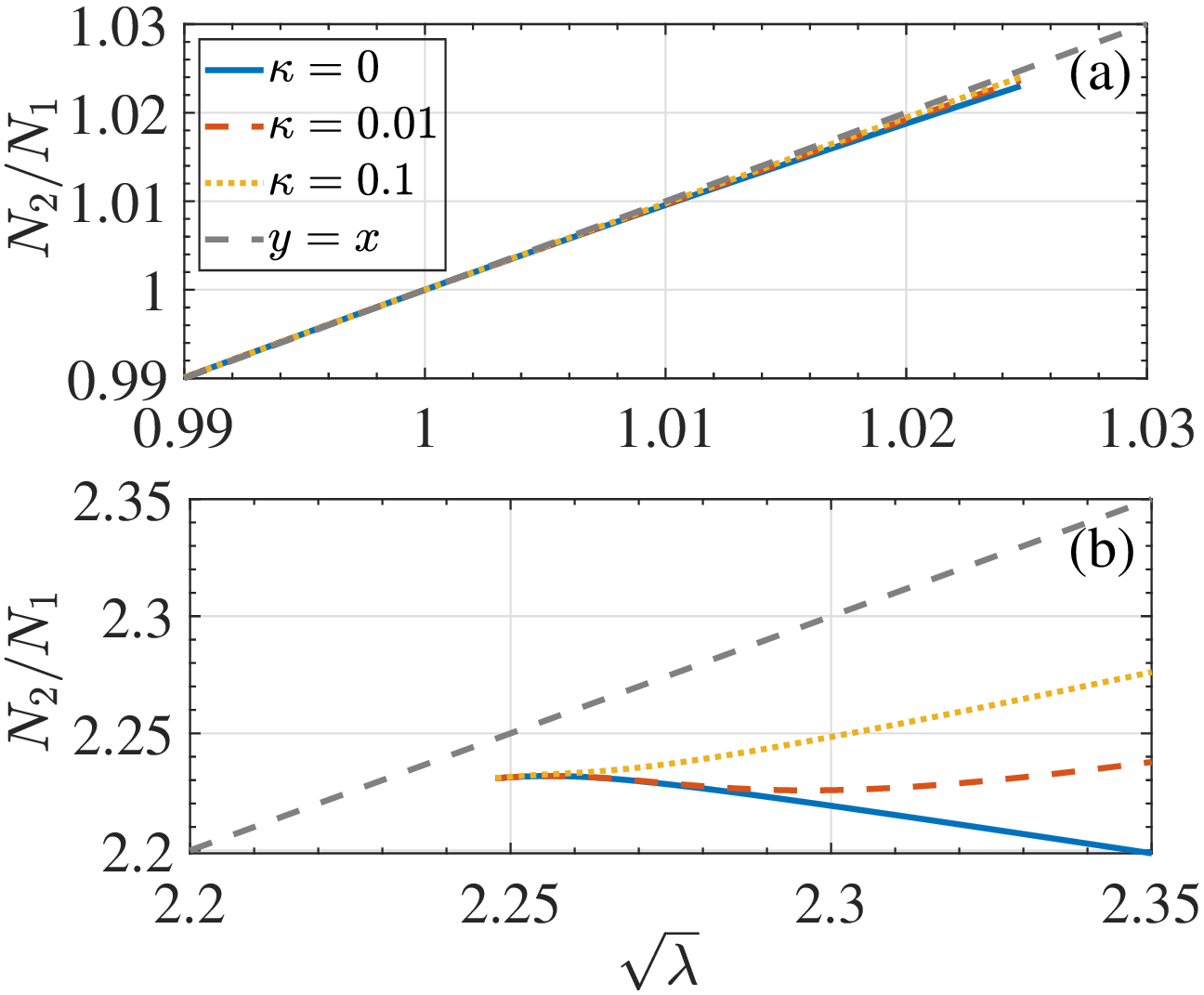}
\caption{\justifying
The ratio of two spin atom numbers or atom number imbalance $N_2/N_1$ in parameter regions (a) I and (b) II, respectively, as a function of the square root of the interaction ratio $\sqrt{\lambda}$ at different values of trapping potential strength $\kappa = 0$ (blue solid lines), $\kappa = 0.01$ (red dashed lines), and $\kappa = 0.1$ (yellow dotted lines), while the dash-dotted line indicates the linear line $y=x$.}
\label{figS:Nratio}
\end{figure}

In addition, Petrov predicted that near the mean-field collapsing point, the density ratio of two spin components in quantum droplets is fixed and associated with the intraspin scattering lengths as $n_2/n_1=\sqrt{a_{11}/a_{22}}$~\cite{petrov2015quantum}, i.e., $n_2/n_1=\sqrt{\lambda}$. By employing a bosonic pairing theory, Hu \emph{et al} showed that this density ratio can vary over a sizable range that could be far away from the fixed ratio $\sqrt{\lambda}$~\cite{hu2025breakdown}. In Fig.~\ref{figS:Nratio}, by numerically solving the extended GPE in Eq.~\eqref{eq:eGPE}, the atom number ratio $N_2/N_1$ is shown as a function of $\sqrt{\lambda}$. In general, the numerical results show that this ratio is not fixed with the variable $\sqrt{\lambda}$ except for the regime when $\sqrt{\lambda}$ is very close to 1. However, the fixed ratio $n_2/n_1=\sqrt{\lambda}$ is adopted in our variational approach and the related calculations, such as density profiles, droplet width, peak density, and collective excitation frequency, show no significant difference with the numerical results.

\bibliography{asymmetricQD1d}

@PREAMBLE{
 "\providecommand{\noopsort}[1]{}" 
 # "\providecommand{\singleletter}[1]{#1}%" 
}

@article{lhy1957eigenvalues,
  title = {Eigenvalues and Eigenfunctions of a Bose System of Hard Spheres and Its Low-Temperature Properties},
  author = {Lee, T. D. and Huang, Kerson and Yang, C. N.},
  journal = {Phys. Rev.},
  volume = {106},
  issue = {6},
  pages = {1135--1145},
  numpages = {0},
  year = {1957},
  month = {Jun},
  publisher = {American Physical Society},
  doi = {10.1103/PhysRev.106.1135},
  url = {https://link.aps.org/doi/10.1103/PhysRev.106.1135}
}

@article{bulgac2002dilute,
  title = {Dilute Quantum Droplets},
  author = {Bulgac, Aurel},
  journal = {Phys. Rev. Lett.},
  volume = {89},
  issue = {5},
  pages = {050402},
  numpages = {4},
  year = {2002},
  month = {Jul},
  publisher = {American Physical Society},
  doi = {10.1103/PhysRevLett.89.050402},
  url = {https://link.aps.org/doi/10.1103/PhysRevLett.89.050402}
}

@article{petrov2015quantum,
  title = {Quantum Mechanical Stabilization of a Collapsing Bose-Bose Mixture},
  author = {Petrov, D. S.},
  journal = {Phys. Rev. Lett.},
  volume = {115},
  issue = {15},
  pages = {155302},
  numpages = {5},
  year = {2015},
  month = {Oct},
  publisher = {American Physical Society},
  doi = {10.1103/PhysRevLett.115.155302},
  url = {https://link.aps.org/doi/10.1103/PhysRevLett.115.155302}
}

@article{petrov2016ultradilute,
  title = {Ultradilute Low-Dimensional Liquids},
  author = {Petrov, D. S. and Astrakharchik, G. E.},
  journal = {Phys. Rev. Lett.},
  volume = {117},
  issue = {10},
  pages = {100401},
  numpages = {5},
  year = {2016},
  month = {Sep},
  publisher = {American Physical Society},
  doi = {10.1103/PhysRevLett.117.100401},
  url = {https://link.aps.org/doi/10.1103/PhysRevLett.117.100401}
}

@article{petrov2023beyond,
  title={Beyond-mean-field effects in mixtures: few-body and many-body aspects},
  author={Petrov, D. S.},
  journal={arXiv preprint arXiv:2312.05336},
  year = {2023},
  doi = {10.48550/arXiv.2312.05336},
  url = {https://doi.org/10.48550/arXiv.2312.05336}
}

@article{kadau2016observing,
  title={Observing the Rosensweig instability of a quantum ferrofluid},
  author={Kadau, Holger and Schmitt, Matthias and Wenzel, Matthias and Wink, Clarissa and Maier, Thomas and Ferrier-Barbut, Igor and Pfau, Tilman},
  journal={Nature},
  volume={530},
  number={7589},
  pages={194--197},
  year={2016},
  publisher={Nature Publishing Group UK London},
  doi = {https://doi.org/10.1038/nature16485},
  URL = { https://doi.org/10.1038/nature16485}
}

@article{ferrier2016observation,
  title={Observation of quantum droplets in a strongly dipolar Bose gas},
  author={Ferrier-Barbut, Igor and Kadau, Holger and Schmitt, Matthias and Wenzel, Matthias and Pfau, Tilman},
  journal={Phys. Rev. Lett.},
  volume={116},
  number={21},
  pages={215301},
  year={2016},
  publisher={APS},
  doi = {10.1103/PhysRevLett.116.215301},
  url = {https://link.aps.org/doi/10.1103/PhysRevLett.116.215301}
}

@article{chomaz2016quantum,
  title = {Quantum-Fluctuation-Driven Crossover from a Dilute Bose-Einstein Condensate to a Macrodroplet in a Dipolar Quantum Fluid},
  author={Chomaz, L and Baier, S and Petter, D and Mark, MJ and W{\"a}chtler, F and Santos, Luis and Ferlaino, F},
  journal={Phys. Rev. X},
  volume={6},
  number={4},
  pages={041039},
  year={2016},
  publisher={APS},
  doi = {10.1103/PhysRevX.6.041039},
  url = {https://link.aps.org/doi/10.1103/PhysRevX.6.041039}
}

@article{schmitt2016self,
  title={Self-bound droplets of a dilute magnetic quantum liquid},
  author={Schmitt, Matthias and Wenzel, Matthias and B{\"o}ttcher, Fabian and Ferrier-Barbut, Igor and Pfau, Tilman},
  journal={Nature},
  volume={539},
  number={7628},
  pages={259--262},
  year={2016},
  publisher={Nature Publishing Group UK London},
  doi={10.1038/nature20126},
  url={https://doi.org/10.1038/nature20126}
}

@article{tanzi2019observation,
  title={Observation of a dipolar quantum gas with metastable supersolid properties},
  author={Tanzi, Luca and Lucioni, Eleonora and Fam{\`a}, Francesca and Catani, Jacopo and Fioretti, Andrea and Gabbanini, Carlo and Bisset, Russell N and Santos, Luis and Modugno, Giovanni},
  journal={Phys. Rev. Lett.},
  volume={122},
  number={13},
  pages={130405},
  year={2019},
  publisher={APS},
  doi = {10.1103/PhysRevLett.122.130405},
  url = {https://link.aps.org/doi/10.1103/PhysRevLett.122.130405}
}

@article{bottcher2019dilute,
  title={Dilute dipolar quantum droplets beyond the extended Gross-Pitaevskii equation},
  author={B{\"o}ttcher, Fabian and Wenzel, Matthias and Schmidt, Jan-Niklas and Guo, Mingyang and Langen, Tim and Ferrier-Barbut, Igor and Pfau, Tilman and Bomb{\'\i}n, Ra{\'u}l and S{\'a}nchez-Baena, Joan and Boronat, Jordi and others},
  journal={Phys. Rev. Res.},
  volume={1},
  number={3},
  pages={033088},
  year={2019},
  publisher={APS},
  doi = {10.1103/PhysRevResearch.1.033088},
  url = {https://link.aps.org/doi/10.1103/PhysRevResearch.1.033088}
}

@article{cabrera2018quantum,
  author = {C. R. Cabrera and L. Tanzi and J. Sanz and B. Naylor and P. Thomas  and P. Cheiney  and L. Tarruell},
  title = {Quantum liquid droplets in a mixture of Bose-Einstein condensates},
  journal = {Science},
  volume = {359},
  number = {6373},
  pages = {301-304},
  year = {2018},
  doi = {10.1126/science.aao5686},
  URL = {https://www.science.org/doi/abs/10.1126/science.aao5686}
}

@article{semeghini2018self,
  title = {Self-Bound Quantum Droplets of Atomic Mixtures in Free Space},
  author = {Semeghini, G. and Ferioli, G. and Masi, L. and Mazzinghi, C. and Wolswijk, L. and Minardi, F. and Modugno, M. and Modugno, G. and Inguscio, M. and Fattori, M.},
  journal = {Phys. Rev. Lett.},
  volume = {120},
  issue = {23},
  pages = {235301},
  numpages = {5},
  year = {2018},
  month = {Jun},
  publisher = {American Physical Society},
  doi = {10.1103/PhysRevLett.120.235301},
  url = {https://link.aps.org/doi/10.1103/PhysRevLett.120.235301}
}

@article{cheiney2018bright,
  title = {Bright Soliton to Quantum Droplet Transition in a Mixture of Bose-Einstein Condensates},
  author = {Cheiney, P. and Cabrera, C. R. and Sanz, J. and Naylor, B. and Tanzi, L. and Tarruell, L.},
  journal = {Phys. Rev. Lett.},
  volume = {120},
  issue = {13},
  pages = {135301},
  numpages = {6},
  year = {2018},
  month = {Mar},
  publisher = {American Physical Society},
  doi = {10.1103/PhysRevLett.120.135301},
  url = {https://link.aps.org/doi/10.1103/PhysRevLett.120.135301}
}

@article{derrico2019observation,
  title = {Observation of quantum droplets in a heteronuclear bosonic mixture},
  author = {D'Errico, C. and Burchianti, A. and Prevedelli, M. and Salasnich, L. and Ancilotto, F. and Modugno, M. and Minardi, F. and Fort, C.},
  journal = {Phys. Rev. Res.},
  volume = {1},
  issue = {3},
  pages = {033155},
  numpages = {8},
  year = {2019},
  month = {Dec},
  publisher = {American Physical Society},
  doi = {10.1103/PhysRevResearch.1.033155},
  url = {https://link.aps.org/doi/10.1103/PhysRevResearch.1.033155}
}

@article{ferioli2019collisions,
  title = {Collisions of Self-Bound Quantum Droplets},
  author = {Ferioli, Giovanni and Semeghini, Giulia and Masi, Leonardo and Giusti, Giovanni and Modugno, Giovanni and Inguscio, Massimo and Gallem\'{\i}, Albert and Recati, Alessio and Fattori, Marco},
  journal = {Phys. Rev. Lett.},
  volume = {122},
  issue = {9},
  pages = {090401},
  numpages = {5},
  year = {2019},
  month = {Mar},
  publisher = {American Physical Society},
  doi = {10.1103/PhysRevLett.122.090401},
  url = {https://link.aps.org/doi/10.1103/PhysRevLett.122.090401}
}

@article{skov2021observation,
  title = {Observation of a Lee-Huang-Yang Fluid},
  author = {Skov, Thomas G. and Skou, Magnus G. and J\o{}rgensen, Nils B. and Arlt, Jan J.},
  journal = {Phys. Rev. Lett.},
  volume = {126},
  issue = {23},
  pages = {230404},
  numpages = {5},
  year = {2021},
  month = {Jun},
  publisher = {American Physical Society},
  doi = {10.1103/PhysRevLett.126.230404},
  url = {https://link.aps.org/doi/10.1103/PhysRevLett.126.230404}
}

@article{guo2021lhy,
  title = {Lee-Huang-Yang effects in the ultracold mixture of $^{23}\mathrm{Na}$ and $^{87}\mathrm{Rb}$ with attractive interspecies interactions},
  author = {Guo, Zhichao and Jia, Fan and Li, Lintao and Ma, Yinfeng and Hutson, Jeremy M. and Cui, Xiaoling and Wang, Dajun},
  journal = {Phys. Rev. Res.},
  volume = {3},
  issue = {3},
  pages = {033247},
  numpages = {8},
  year = {2021},
  month = {Sep},
  publisher = {American Physical Society},
  doi = {10.1103/PhysRevResearch.3.033247},
  url = {https://link.aps.org/doi/10.1103/PhysRevResearch.3.033247}
}

@article{cavicchioli2025dynamical,
  title = {Dynamical Formation of Multiple Quantum Droplets in a Bose-Bose Mixture},
  author = {Cavicchioli, L. and Fort, C. and Ancilotto, F. and Modugno, M. and Minardi, F. and Burchianti, A.},
  journal = {Phys. Rev. Lett.},
  volume = {134},
  issue = {9},
  pages = {093401},
  numpages = {6},
  year = {2025},
  month = {Mar},
  publisher = {American Physical Society},
  doi = {10.1103/PhysRevLett.134.093401},
  url = {https://link.aps.org/doi/10.1103/PhysRevLett.134.093401}
}

@Article{debraj2019quantum,
	title={{Quantum Bose-Fermi droplets}},
	author={Debraj Rakshit and Tomasz Karpiuk and Mirosław Brewczyk and Mariusz Gajda},
	journal={SciPost Phys.},
	volume={6},
	pages={079},
	year={2019},
	publisher={SciPost},
	doi={10.21468/SciPostPhys.6.6.079},
	url={https://scipost.org/10.21468/SciPostPhys.6.6.079},
}

@article{rakshit2019Sselfbound,
doi = {10.1088/1367-2630/ab2ce3},
url = {https://dx.doi.org/10.1088/1367-2630/ab2ce3},
year = {2019},
month = {jul},
publisher = {IOP Publishing},
volume = {21},
number = {7},
pages = {073027},
author = {Rakshit, Debraj and Karpiuk, Tomasz and Zin, Paweł and Brewczyk, Mirosław and Lewenstein, Maciej and Gajda, Mariusz},
title = {Self-bound Bose–Fermi liquids in lower dimensions},
journal = {New Journal of Physics}
}

@article{smith2021quantum,
  title = {Quantum Droplet States of a Binary Magnetic Gas},
  author = {Smith, Joseph C. and Baillie, D. and Blakie, P. B.},
  journal = {Phys. Rev. Lett.},
  volume = {126},
  issue = {2},
  pages = {025302},
  numpages = {6},
  year = {2021},
  month = {Jan},
  publisher = {American Physical Society},
  doi = {10.1103/PhysRevLett.126.025302},
  url = {https://link.aps.org/doi/10.1103/PhysRevLett.126.025302}
}

@article{bisset2021quantum,
  title = {Quantum Droplets of Dipolar Mixtures},
  author = {Bisset, R. N. and Ardila, L. A. Pe\~na and Santos, L.},
  journal = {Phys. Rev. Lett.},
  volume = {126},
  issue = {2},
  pages = {025301},
  numpages = {6},
  year = {2021},
  month = {Jan},
  publisher = {American Physical Society},
  doi = {10.1103/PhysRevLett.126.025301},
  url = {https://link.aps.org/doi/10.1103/PhysRevLett.126.025301}
}

@article{kartashov2019frontiers,
  title={Frontiers in multidimensional self-trapping of nonlinear fields and matter},
  author={Kartashov, Yaroslav V and Astrakharchik, Gregory E and Malomed, Boris A and Torner, Lluis},
  journal={Nat. Rev. Phys.},
  volume={1},
  number={3},
  pages={185--197},
  year={2019},
  publisher={Nature Publishing Group UK London},
  doi = {10.1038/s42254-019-0025-7},
  url = {https://doi.org/10.1038/s42254-019-0025-7}
}

@article{bottcher2021new,
  title={New states of matter with fine-tuned interactions: quantum droplets and dipolar supersolids},
  author={B{\"o}ttcher, Fabian and Schmidt, Jan-Niklas and Hertkorn, Jens and Ng, Kevin SH and Graham, Sean D and Guo, Mingyang and Langen, Tim and Pfau, Tilman},
  journal={Rep. Prog. Phys.},
  volume={84},
  number={1},
  pages={012403},
  year={2021},
  publisher={IOP Publishing},
  doi = {10.1088/1361-6633/abc9ab},
  url = {https://dx.doi.org/10.1088/1361-6633/abc9ab}
}

@article{luo2021new,
  title={A new form of liquid matter: Quantum droplets},
  author={Luo, Zhi-Huan and Pang, Wei and Liu, Bin and Li, Yong-Yao and Malomed, Boris A},
  journal={Frontiers of Physics},
  volume={16},
  number={3},
  pages={32201},
  year={2021},
  publisher={Springer},
  doi = {10.1007/s11467-020-1020-2},
  url = {https://doi.org/10.1007/s11467-020-1020-2}
}

@article{guo2021new,
  title={A new state of matter of quantum droplets},
  author={Guo, Mingyang and Pfau, Tilman},
  journal={Frontiers of Physics},
  volume={16},
  number={3},
  pages={32202},
  year={2021},
  publisher={Springer},
  doi = {10.1007/s11467-020-1035-8},
  url = {https://doi.org/10.1007/s11467-020-1035-8}
}

@article{cikojevic2018ultradilute,
  title = {Ultradilute quantum liquid drops},
  author = {Cikojevi{\'c}, Viktor and D{\v{z}}elalija, Kre{\v{s}}imir and Stipanovi{\'c}, Petar and Vranje{\v{s}} Marki{\'c}, Leandra and Boronat, Jordi},
  journal = {Phys. Rev. B},
  volume = {97},
  issue = {14},
  pages = {140502},
  numpages = {5},
  year = {2018},
  month = {Apr},
  publisher = {American Physical Society},
  doi = {10.1103/PhysRevB.97.140502},
  url = {https://link.aps.org/doi/10.1103/PhysRevB.97.140502}
}

@article{parisi2019liquid,
  title={Liquid state of one-dimensional Bose mixtures: a quantum Monte Carlo study},
  author={Parisi, L and Astrakharchik, GE and Giorgini, Stefano},
  journal={Phys. Rev. Lett.},
  volume={122},
  number={10},
  pages={105302},
  year={2019},
  publisher={APS},
  doi = {10.1103/PhysRevLett.122.105302},
  url = {https://link.aps.org/doi/10.1103/PhysRevLett.122.105302}
}

@article{staudinger2018selfbound,
  title = {Self-bound Bose mixtures},
  author = {Staudinger, Clemens and Mazzanti, Ferran and Zillich, Robert E.},
  journal = {Phys. Rev. A},
  volume = {98},
  issue = {2},
  pages = {023633},
  numpages = {5},
  year = {2018},
  month = {Aug},
  publisher = {American Physical Society},
  doi = {10.1103/PhysRevA.98.023633},
  url = {https://link.aps.org/doi/10.1103/PhysRevA.98.023633}
}

@article{hu2020consistent,
  title = {Consistent Theory of Self-Bound Quantum Droplets with Bosonic Pairing},
  author = {Hu, Hui and Liu, Xia-Ji},
  journal = {Phys. Rev. Lett.},
  volume = {125},
  issue = {19},
  pages = {195302},
  numpages = {6},
  year = {2020},
  month = {Nov},
  publisher = {American Physical Society},
  doi = {10.1103/PhysRevLett.125.195302},
  url = {https://link.aps.org/doi/10.1103/PhysRevLett.125.195302}
}

@article{gu2020phonon,
  title = {Phonon stability and sound velocity of quantum droplets in a boson mixture},
  author = {Gu, Qi and Yin, Lan},
  journal = {Phys. Rev. B},
  volume = {102},
  issue = {22},
  pages = {220503},
  numpages = {5},
  year = {2020},
  month = {Dec},
  publisher = {American Physical Society},
  doi = {10.1103/PhysRevB.102.220503},
  url = {https://link.aps.org/doi/10.1103/PhysRevB.102.220503}
}

@article{pan2022quantum,
  title = {Quantum phases of self-bound droplets of Bose-Bose mixtures},
  author = {Pan, Junqiao and Yi, Su and Shi, Tao},
  journal = {Phys. Rev. Res.},
  volume = {4},
  issue = {4},
  pages = {043018},
  numpages = {15},
  year = {2022},
  month = {Oct},
  publisher = {American Physical Society},
  doi = {10.1103/PhysRevResearch.4.043018},
  url = {https://link.aps.org/doi/10.1103/PhysRevResearch.4.043018}
}

@article{he2023quantum,
  title = {Quantum Criticality of Liquid-Gas Transition in a Binary Bose Mixture},
  author = {He, Li and Li, Haowei and Yi, Wei and Yu, Zeng-Qiang},
  journal = {Phys. Rev. Lett.},
  volume = {130},
  issue = {19},
  pages = {193001},
  numpages = {7},
  year = {2023},
  month = {May},
  publisher = {American Physical Society},
  doi = {10.1103/PhysRevLett.130.193001},
  url = {https://link.aps.org/doi/10.1103/PhysRevLett.130.193001}
}

@article{cappellaro2018collective,
  title = {Collective modes across the soliton-droplet crossover in binary Bose mixtures},
  author = {Cappellaro, Alberto and Macr\`{\i}, Tommaso and Salasnich, Luca},
  journal = {Phys. Rev. A},
  volume = {97},
  issue = {5},
  pages = {053623},
  numpages = {7},
  year = {2018},
  month = {May},
  publisher = {American Physical Society},
  doi = {10.1103/PhysRevA.97.053623},
  url = {https://link.aps.org/doi/10.1103/PhysRevA.97.053623}
}

@article{spada2023attractive,
  title = {Attractive Solution of Binary Bose Mixtures: Liquid-Vapor Coexistence and Critical Point},
  author = {Spada, G. and Pilati, S. and Giorgini, S.},
  journal = {Phys. Rev. Lett.},
  volume = {131},
  issue = {17},
  pages = {173404},
  numpages = {6},
  year = {2023},
  month = {Oct},
  publisher = {American Physical Society},
  doi = {10.1103/PhysRevLett.131.173404},
  url = {https://link.aps.org/doi/10.1103/PhysRevLett.131.173404}
}

@article{kartashov2018three,
  title={Three-dimensional droplets of swirling superfluids},
  author={Kartashov, Yaroslav V and Malomed, Boris A and Tarruell, Leticia and Torner, Lluis},
  journal={Phys. Rev. A},
  volume={98},
  number={1},
  pages={013612},
  year={2018},
  publisher={APS},
  doi = {10.1103/PhysRevA.98.013612},
  url = {https://link.aps.org/doi/10.1103/PhysRevA.98.013612}
}

@article{li2018two,
  title = {Two-dimensional vortex quantum droplets},
  author = {Li, Yongyao and Chen, Zhaopin and Luo, Zhihuan and Huang, Chunqing and Tan, Haishu and Pang, Wei and Malomed, Boris A.},
  journal = {Phys. Rev. A},
  volume = {98},
  issue = {6},
  pages = {063602},
  numpages = {12},
  year = {2018},
  month = {Dec},
  publisher = {American Physical Society},
  doi = {10.1103/PhysRevA.98.063602},
  url = {https://link.aps.org/doi/10.1103/PhysRevA.98.063602}
}

@article{kartashov2019metastability,
  title = {Metastability of Quantum Droplet Clusters},
  author = {Kartashov, Yaroslav V. and Malomed, Boris A. and Torner, Lluis},
  journal = {Phys. Rev. Lett.},
  volume = {122},
  issue = {19},
  pages = {193902},
  numpages = {7},
  year = {2019},
  month = {May},
  publisher = {American Physical Society},
  doi = {10.1103/PhysRevLett.122.193902},
  url = {https://link.aps.org/doi/10.1103/PhysRevLett.122.193902}
}

@article{otajonov2020variational,
  title = {Variational approximation for two-dimensional quantum droplets},
  author = {Otajonov, Sherzod R. and Tsoy, Eduard N. and Abdullaev, Fatkhulla Kh.},
  journal = {Phys. Rev. E},
  volume = {102},
  issue = {6},
  pages = {062217},
  numpages = {7},
  year = {2020},
  month = {Dec},
  publisher = {American Physical Society},
  doi = {10.1103/PhysRevE.102.062217},
  url = {https://link.aps.org/doi/10.1103/PhysRevE.102.062217}
}

@article{guebli2021quantum,
  title = {Quantum self-bound droplets in Bose-Bose mixtures: Effects of higher-order quantum and thermal fluctuations},
  author = {Guebli, Nadia and Boudjem\^aa, Abdel\^aali},
  journal = {Phys. Rev. A},
  volume = {104},
  issue = {2},
  pages = {023310},
  numpages = {6},
  year = {2021},
  month = {Aug},
  publisher = {American Physical Society},
  doi = {10.1103/PhysRevA.104.023310},
  url = {https://link.aps.org/doi/10.1103/PhysRevA.104.023310}
}

@article{zin2018quantum,
  title = {Quantum Bose-Bose droplets at a dimensional crossover},
  author = {Zin, Pawe\l{} and Pylak, Maciej and Wasak, Tomasz and Gajda, Mariusz and Idziaszek, Zbigniew},
  journal = {Phys. Rev. A},
  volume = {98},
  issue = {5},
  pages = {051603},
  numpages = {5},
  year = {2018},
  month = {Nov},
  publisher = {American Physical Society},
  doi = {10.1103/PhysRevA.98.051603},
  url = {https://link.aps.org/doi/10.1103/PhysRevA.98.051603}
}

@article{cikojevic2021dynamics,
  title = {Dynamics of equilibration and collisions in ultradilute quantum droplets},
  author = {Cikojevi\ifmmode \acute{c}\else \'{c}\fi{}, V. and Marki\ifmmode \acute{c}\else \'{c}\fi{}, L. Vranje\ifmmode \check{s}\else \v{s}\fi{} and Pi, M. and Barranco, M. and Ancilotto, F. and Boronat, J.},
  journal = {Phys. Rev. Res.},
  volume = {3},
  issue = {4},
  pages = {043139},
  numpages = {11},
  year = {2021},
  month = {Nov},
  publisher = {American Physical Society},
  doi = {10.1103/PhysRevResearch.3.043139},
  url = {https://link.aps.org/doi/10.1103/PhysRevResearch.3.043139}
}

@article{mithun2021statistical,
  title={Statistical mechanics of one-dimensional quantum droplets},
  author={Mithun, T and Mistakidis, SI and Schmelcher, P and Kevrekidis, PG},
  journal={Phys. Rev. A},
  volume={104},
  number={3},
  pages={033316},
  year={2021},
  doi = {10.1103/PhysRevA.104.033316},
  url = {https://link.aps.org/doi/10.1103/PhysRevA.104.033316},
  publisher={APS}
}

@Article{hu2022collisional,
author={Hu, Yanming
and Fei, Yifan
and Chen, XiaoLong
and Zhang, Yunbo},
title={Collisional dynamics of symmetric two-dimensional quantum droplets},
journal={Front. Phys.},
year={2022},
month={Aug},
day={19},
volume={17},
number={6},
pages={61505},
issn={2095-0470},
doi={10.1007/s11467-022-1192-z},
url={https://doi.org/10.1007/s11467-022-1192-z}
}

@article{tengstrand2022droplet,
  title={Droplet-superfluid compounds in binary bosonic mixtures},
  author={Tengstrand, M Nilsson and Reimann, SM},
  journal={Phys. Rev. A},
  volume={105},
  number={3},
  pages={033319},
  year={2022},
  publisher={APS},
  doi = {10.1103/PhysRevA.105.033319},
  url = {https://link.aps.org/doi/10.1103/PhysRevA.105.033319}
}

@article{nikolaou2023rotating,
  title={Rotating quantum droplets confined in a harmonic potential},
  author={Nikolaou, S and Kavoulakis, GM and {\"O}gren, M},
  journal={Phys. Rev. A},
  volume={108},
  number={5},
  pages={053309},
  year={2023},
  publisher={APS},
  doi = {10.1103/PhysRevA.108.053309},
  url = {https://link.aps.org/doi/10.1103/PhysRevA.108.053309}
}

@article{astrakharchik2018dynamics,
  title = {Dynamics of one-dimensional quantum droplets},
  author = {Astrakharchik, G. E. and Malomed, B. A.},
  journal = {Phys. Rev. A},
  volume = {98},
  issue = {1},
  pages = {013631},
  numpages = {11},
  year = {2018},
  month = {Jul},
  publisher = {American Physical Society},
  doi = {10.1103/PhysRevA.98.013631},
  url = {https://link.aps.org/doi/10.1103/PhysRevA.98.013631}
}

@article{otajonov2019stationary,
title = {Stationary and dynamical properties of one-dimensional quantum droplets},
journal = {Physics Letters A},
volume = {383},
number = {34},
pages = {125980},
year = {2019},
issn = {0375-9601},
doi = {https://doi.org/10.1016/j.physleta.2019.125980},
url = {https://www.sciencedirect.com/science/article/pii/S0375960119308473},
author = {Sherzod R. Otajonov and Eduard N. Tsoy and Fatkhulla Kh. Abdullaev}
}

@article{sturmer2021breathing,
  title = {Breathing mode in two-dimensional binary self-bound Bose-gas droplets},
  author = {St\"urmer, P. and Tengstrand, M. Nilsson and Sachdeva, R. and Reimann, S. M.},
  journal = {Phys. Rev. A},
  volume = {103},
  issue = {5},
  pages = {053302},
  numpages = {9},
  year = {2021},
  month = {May},
  publisher = {American Physical Society},
  doi = {10.1103/PhysRevA.103.053302},
  url = {https://link.aps.org/doi/10.1103/PhysRevA.103.053302}
}

@article{tylutki2020collective,
  title = {Collective excitations of a one-dimensional quantum droplet},
  author = {Tylutki, Marek and Astrakharchik, Grigori E. and Malomed, Boris A. and Petrov, Dmitry S.},
  journal = {Phys. Rev. A},
  volume = {101},
  issue = {5},
  pages = {051601},
  numpages = {5},
  year = {2020},
  month = {May},
  publisher = {American Physical Society},
  doi = {10.1103/PhysRevA.101.051601},
  url = {https://link.aps.org/doi/10.1103/PhysRevA.101.051601}
}

@article{hu2020collective,
  title = {Collective excitations of a spherical ultradilute quantum droplet},
  author = {Hu, Hui and Liu, Xia-Ji},
  journal = {Phys. Rev. A},
  volume = {102},
  issue = {5},
  pages = {053303},
  numpages = {16},
  year = {2020},
  month = {Nov},
  publisher = {American Physical Society},
  doi = {10.1103/PhysRevA.102.053303},
  url = {https://link.aps.org/doi/10.1103/PhysRevA.102.053303}
}

@article{dong2022internal,
  title = {Internal modes of two-dimensional quantum droplets},
  author = {Dong, Liangwei and Shi, Kai and Huang, Changming},
  journal = {Phys. Rev. A},
  volume = {106},
  issue = {5},
  pages = {053303},
  numpages = {7},
  year = {2022},
  month = {Nov},
  publisher = {American Physical Society},
  doi = {10.1103/PhysRevA.106.053303},
  url = {https://link.aps.org/doi/10.1103/PhysRevA.106.053303}
}

@article{du2023ground,
  title = {Ground-state properties and Bogoliubov modes of a harmonically trapped one-dimensional quantum droplet},
  author = {Du, Xucong and Fei, Yifan and Chen, Xiao-Long and Zhang, Yunbo},
  journal = {Phys. Rev. A},
  volume = {108},
  issue = {3},
  pages = {033312},
  numpages = {10},
  year = {2023},
  month = {Sep},
  publisher = {American Physical Society},
  doi = {10.1103/PhysRevA.108.033312},
  url = {https://link.aps.org/doi/10.1103/PhysRevA.108.033312}
}

@article{fei2024collective,
  title = {Collective excitations in two-dimensional harmonically trapped quantum droplets},
  author = {Fei, Yifan and Du, Xucong and Chen, Xiao-Long and Zhang, Yunbo},
  journal = {Phys. Rev. A},
  volume = {109},
  issue = {5},
  pages = {053309},
  numpages = {11},
  year = {2024},
  month = {May},
  publisher = {American Physical Society},
  doi = {10.1103/PhysRevA.109.053309},
  url = {https://link.aps.org/doi/10.1103/PhysRevA.109.053309}
}

@article{charalampidis2025twocomponent,
  title = {Two-component droplet phases and their spectral stability in one dimension},
  author = {Charalampidis, E. G. and Mistakidis, S. I.},
  journal = {Phys. Rev. A},
  volume = {111},
  issue = {1},
  pages = {013318},
  numpages = {12},
  year = {2025},
  month = {Jan},
  publisher = {American Physical Society},
  doi = {10.1103/PhysRevA.111.013318},
  url = {https://link.aps.org/doi/10.1103/PhysRevA.111.013318}
}

@article{flynn2023quantum,
  title = {Quantum droplets in imbalanced atomic mixtures},
  author = {Flynn, T. A. and Parisi, L. and Billam, T. P. and Parker, N. G.},
  journal = {Phys. Rev. Res.},
  volume = {5},
  issue = {3},
  pages = {033167},
  numpages = {12},
  year = {2023},
  month = {Sep},
  publisher = {American Physical Society},
  doi = {10.1103/PhysRevResearch.5.033167},
  url = {https://link.aps.org/doi/10.1103/PhysRevResearch.5.033167}
}

@article{flynn2024harmonically,
  title = {Harmonically trapped imbalanced quantum droplets},
  author = {Flynn, T. A. and Keepfer, N. A. and Parker, N. G. and Billam, T. P.},
  journal = {Phys. Rev. Res.},
  volume = {6},
  issue = {1},
  pages = {013209},
  numpages = {11},
  year = {2024},
  month = {Feb},
  publisher = {American Physical Society},
  doi = {10.1103/PhysRevResearch.6.013209},
  url = {https://link.aps.org/doi/10.1103/PhysRevResearch.6.013209}
}

@article{kartashov2024multipole,
  title = {Multipole quantum droplets in quasi-one-dimensional asymmetric mixtures},
  author = {Kartashov, Yaroslav V. and Zezyulin, Dmitry A.},
  journal = {Phys. Rev. A},
  volume = {110},
  issue = {2},
  pages = {L021304},
  numpages = {7},
  year = {2024},
  month = {Aug},
  publisher = {American Physical Society},
  doi = {10.1103/PhysRevA.110.L021304},
  url = {https://link.aps.org/doi/10.1103/PhysRevA.110.L021304}
}

@article{li2012sumrules,
doi = {10.1209/0295-5075/99/56008},
url = {https://dx.doi.org/10.1209/0295-5075/99/56008},
year = {2012},
month = {sep},
publisher = {EDP Sciences, IOP Publishing and Società Italiana di Fisica},
volume = {99},
number = {5},
pages = {56008},
author = {Li, Yun and Martone, Giovanni Italo and Stringari, Sandro},
title = {Sum rules, dipole oscillation and spin polarizability of a spin-orbit coupled quantum gas},
journal = {Europhysics Letters}
}

@article{sartori2015spindipole,
doi = {10.1088/1367-2630/17/9/093036},
url = {https://dx.doi.org/10.1088/1367-2630/17/9/093036},
year = {2015},
month = {sep},
publisher = {IOP Publishing},
volume = {17},
number = {9},
pages = {093036},
author = {Sartori, A and Marino, J and Stringari, S and Recati, A},
title = {Spin-dipole oscillation and relaxation of coherently coupled Bose–Einstein condensates},
journal = {New Journal of Physics}
}

@article{bienaim2016soc,
  title = {Spin-dipole oscillation and polarizability of a binary Bose-Einstein condensate near the miscible-immiscible phase transition},
  author = {Bienaim\'e, Tom and Fava, Eleonora and Colzi, Giacomo and Mordini, Carmelo and Serafini, Simone and Qu, Chunlei and Stringari, Sandro and Lamporesi, Giacomo and Ferrari, Gabriele},
  journal = {Phys. Rev. A},
  volume = {94},
  issue = {6},
  pages = {063652},
  numpages = {5},
  year = {2016},
  month = {Dec},
  publisher = {American Physical Society},
  doi = {10.1103/PhysRevA.94.063652},
  url = {https://link.aps.org/doi/10.1103/PhysRevA.94.063652}
}

@article{fava2018observation,
  title = {Observation of Spin Superfluidity in a Bose Gas Mixture},
  author = {Fava, Eleonora and Bienaim\'e, Tom and Mordini, Carmelo and Colzi, Giacomo and Qu, Chunlei and Stringari, Sandro and Lamporesi, Giacomo and Ferrari, Gabriele},
  journal = {Phys. Rev. Lett.},
  volume = {120},
  issue = {17},
  pages = {170401},
  numpages = {6},
  year = {2018},
  month = {Apr},
  publisher = {American Physical Society},
  doi = {10.1103/PhysRevLett.120.170401},
  url = {https://link.aps.org/doi/10.1103/PhysRevLett.120.170401}
}

@article{mithun2020modulational,
  AUTHOR = {Mithun, Thudiyangal and Maluckov, Aleksandra and Kasamatsu, Kenichi and Malomed, Boris A. and Khare, Avinash},
  TITLE = {Modulational Instability, Inter-Component Asymmetry, and Formation of Quantum Droplets in One-Dimensional Binary Bose Gases},
  JOURNAL = {Symmetry},
  VOLUME = {12},
  YEAR = {2020},
  NUMBER = {1},
  pages = {174},
  URL = {https://www.mdpi.com/2073-8994/12/1/174},
  ISSN = {2073-8994},
  DOI = {10.3390/sym12010174}
}

@article{stringari1996collective,
  title = {Collective Excitations of a Trapped Bose-Condensed Gas},
  author = {Stringari, S.},
  journal = {Phys. Rev. Lett.},
  volume = {77},
  issue = {12},
  pages = {2360--2363},
  numpages = {0},
  year = {1996},
  month = {Sep},
  publisher = {American Physical Society},
  doi = {10.1103/PhysRevLett.77.2360},
  url = {https://link.aps.org/doi/10.1103/PhysRevLett.77.2360}
}

@article{menotti2002collective,
  title = {Collective oscillations of a one-dimensional trapped Bose-Einstein gas},
  author = {Menotti, Chiara and Stringari, Sandro},
  journal = {Phys. Rev. A},
  volume = {66},
  issue = {4},
  pages = {043610},
  numpages = {6},
  year = {2002},
  month = {Oct},
  publisher = {American Physical Society},
  doi = {10.1103/PhysRevA.66.043610},
  url = {https://link.aps.org/doi/10.1103/PhysRevA.66.043610}
}

@article{dalfovo1999theory,
  title = {Theory of Bose-Einstein condensation in trapped gases},
  author = {Dalfovo, Franco and Giorgini, Stefano and Pitaevskii, Lev P. and Stringari, Sandro},
  journal = {Rev. Mod. Phys.},
  volume = {71},
  issue = {3},
  pages = {463--512},
  numpages = {0},
  year = {1999},
  month = {Apr},
  publisher = {American Physical Society},
  doi = {10.1103/RevModPhys.71.463},
  url = {https://link.aps.org/doi/10.1103/RevModPhys.71.463}
}

@book{pitaevskii2016bose,
  title={Bose-Einstein condensation and superfluidity},
  author={Pitaevskii, Lev and Stringari, Sandro},
  volume={164},
  year={2016},
  publisher={Oxford University Press},
  doi={10.1093/acprof:oso/9780198758884.001.0001},
  url = {https://doi.org/10.1093/acprof:oso/9780198758884.001.0001}
}

@article{gorlitz2001realization,
  title = {Realization of Bose-Einstein Condensates in Lower Dimensions},
  author = {G\"orlitz, A. and Vogels, J. M. and Leanhardt, A. E. and Raman, C. and Gustavson, T. L. and Abo-Shaeer, J. R. and Chikkatur, A. P. and Gupta, S. and Inouye, S. and Rosenband, T. and Ketterle, W.},
  journal = {Phys. Rev. Lett.},
  volume = {87},
  issue = {13},
  pages = {130402},
  numpages = {4},
  year = {2001},
  month = {Sep},
  publisher = {American Physical Society},
  doi = {10.1103/PhysRevLett.87.130402},
  url = {https://link.aps.org/doi/10.1103/PhysRevLett.87.130402}
}

@article{schreck2001quasipure,
  title = {Quasipure Bose-Einstein Condensate Immersed in a Fermi Sea},
  author = {Schreck, F. and Khaykovich, L. and Corwin, K. L. and Ferrari, G. and Bourdel, T. and Cubizolles, J. and Salomon, C.},
  journal = {Phys. Rev. Lett.},
  volume = {87},
  issue = {8},
  pages = {080403},
  numpages = {4},
  year = {2001},
  month = {Aug},
  publisher = {American Physical Society},
  doi = {10.1103/PhysRevLett.87.080403},
  url = {https://link.aps.org/doi/10.1103/PhysRevLett.87.080403}
}

@article{greiner2001exploring,
  title = {Exploring Phase Coherence in a 2D Lattice of Bose-Einstein Condensates},
  author = {Greiner, Markus and Bloch, Immanuel and Mandel, Olaf and H\"ansch, Theodor W. and Esslinger, Tilman},
  journal = {Phys. Rev. Lett.},
  volume = {87},
  issue = {16},
  pages = {160405},
  numpages = {4},
  year = {2001},
  month = {Oct},
  publisher = {American Physical Society},
  doi = {10.1103/PhysRevLett.87.160405},
  url = {https://link.aps.org/doi/10.1103/PhysRevLett.87.160405}
}

@article{olshanii1998atomic,
  title = {Atomic Scattering in the Presence of an External Confinement and a Gas of Impenetrable Bosons},
  author = {Olshanii, M.},
  journal = {Phys. Rev. Lett.},
  volume = {81},
  issue = {5},
  pages = {938--941},
  numpages = {0},
  year = {1998},
  month = {Aug},
  publisher = {American Physical Society},
  doi = {10.1103/PhysRevLett.81.938},
  url = {https://link.aps.org/doi/10.1103/PhysRevLett.81.938}
}

@article{derrico2007feshbach,
doi = {10.1088/1367-2630/9/7/223},
url = {https://dx.doi.org/10.1088/1367-2630/9/7/223},
year = {2007},
month = {jul},
publisher = {},
volume = {9},
number = {7},
pages = {223},
author = {Chiara D'Errico and Matteo Zaccanti and Marco Fattori and Giacomo Roati and Massimo Inguscio and Giovanni Modugno and Andrea Simoni},
title = {Feshbach resonances in ultracold 39K},
journal = {New Journal of Physics}
}

@article{parisi2020quantum,
  title={Quantum droplets in one-dimensional Bose mixtures: A quantum Monte Carlo study},
  author={Parisi, Luca and Giorgini, Stefano},
  journal={Phys. Rev. A},
  volume={102},
  number={2},
  pages={023318},
  year={2020},
  publisher={APS},
  doi = {10.1103/PhysRevA.102.023318},
  url = {https://link.aps.org/doi/10.1103/PhysRevA.102.023318}
}

@article{mistakidis2021formation,
  title={Formation and quench of homonuclear and heteronuclear quantum droplets in one dimension},
  author={Mistakidis, SI and Mithun, T and Kevrekidis, PG and Sadeghpour, HR and Schmelcher, Peter},
  journal={Phys. Rev. Res.},
  volume={3},
  number={4},
  pages={043128},
  year={2021},
  publisher={APS},
  doi = {10.1103/PhysRevResearch.3.043128},
  url = {https://link.aps.org/doi/10.1103/PhysRevResearch.3.043128}
}

@article{kimura2002Breathing,
  title = {Breathing modes of Bose-Einstein condensates in highly asymmetric traps},
  author = {Kimura, Takashi},
  journal = {Phys. Rev. A},
  volume = {66},
  issue = {1},
  pages = {013608},
  numpages = {5},
  year = {2002},
  month = {Jul},
  publisher = {American Physical Society},
  doi = {10.1103/PhysRevA.66.013608},
  url = {https://link.aps.org/doi/10.1103/PhysRevA.66.013608}
}

@article{geier2021exciting,
  title = {Exciting the Goldstone Modes of a Supersolid Spin-Orbit-Coupled Bose Gas},
  author = {Geier, Kevin T. and Martone, Giovanni I. and Hauke, Philipp and Stringari, Sandro},
  journal = {Phys. Rev. Lett.},
  volume = {127},
  issue = {11},
  pages = {115301},
  numpages = {6},
  year = {2021},
  month = {Sep},
  publisher = {American Physical Society},
  doi = {10.1103/PhysRevLett.127.115301},
  url = {https://link.aps.org/doi/10.1103/PhysRevLett.127.115301}
}

@misc{ritu2026spin,
      title={Spin and density excitations of one-dimensional self-bound Bose-Bose droplets}, 
      author={Ritu and Rajat and Manpreet Singh and Rajesh Kumar Gupta and Sandeep Gautam},
      year={2026},
      eprint={2603.00987},
      archivePrefix={arXiv},
      primaryClass={cond-mat.quant-gas},
      url={https://arxiv.org/abs/2603.00987}, 
}

@article{hu2025breakdown,
  title = {Breakdown of the single-mode description of ultradilute quantum droplets in binary Bose mixtures: A perspective from a microscopic bosonic pairing theory},
  author = {Hu, Hui and Wang, Jia and Pu, Han and Liu, Xia-Ji},
  journal = {Phys. Rev. A},
  volume = {111},
  issue = {2},
  pages = {023309},
  numpages = {11},
  year = {2025},
  month = {Feb},
  publisher = {American Physical Society},
  doi = {10.1103/PhysRevA.111.023309},
  url = {https://link.aps.org/doi/10.1103/PhysRevA.111.023309}
}

@dataset{xiao2026data,
  author       = {Xiao, Huiyun and Zhang, Xinran and Liu, Junli and Du, Xucong and Chen, Xiao-Long and Zhang, Yunbo},
  title        = {Data for One-dimensional asymmetrically interacting quantum droplets in Bose-Bose mixtures},
  month        = mar,
  year         = 2026,
  publisher    = {Zenodo},
  doi          = {10.5281/zenodo.19206410},
  url          = {https://doi.org/10.5281/zenodo.19206410},
}

\end{document}